\documentclass[apjl,iop]{emulateapj}

\usepackage{amssymb,natbib,graphicx,subfigure,color,apjfonts}

\def\AREPO{{\small AREPO}}

\def\FOF{{\small FOF}}
\def\SUBFIND{{\small SUBFIND}}
\def\SUBLINK{{\small SUBLINK}}

\def\log{{\rm\thinspace log}}

\newcommand{\ifm}[1]{\relax\ifmmode#1\else$\mathsurround=0pt #1$\fi}
\newcommand{\be}{\begin{equation}}
\newcommand{\ee}{\end{equation}}
\newcommand{\bea}{\begin{eqnarray}}
\newcommand{\eea}{\end{eqnarray}}
\newcommand{\equ}[1]{equation (\ref{e:#1})}
\newcommand{\Fig}[1]{Fig.~\ref{f:#1}}
\newcommand{\Figs}[2]{Figs.~\ref{f:#1} and \ref{f:#2}}
\newcommand{\sggg}[1]{\textcolor{green}{[]}}

\def\dex{{\rm\thinspace dex}}

\def\pc{{\rm\thinspace pc}}
\def\kpc{{\rm\thinspace kpc}}
\def\Mpc{{\rm\thinspace Mpc}}

\def\pcmcb{{\rm\thinspace cm}^{-3}}

\def\erg{{\rm\thinspace erg}}

\def\Msun{\hbox{$\rm\thinspace M_{\odot}$}}

\def\yr{{\rm\thinspace yr}}
\def\Myr{{\rm\thinspace Myr}}

\def\Msunpc2{{\Msun\pc}^{-2}}
\def\Msunyrkpc2{{\Msun\yr^{-1}\kpc}^{-2}}

\def\magarcsec2{{\rm\thinspace mag\thinspace arcsec}^{-2}}

\shorttitle{Tracing angular momentum in Illustris}
\shorttitle{Lagrangian angular momentum in Illustris}
\shorttitle{Impact of galactic wind on angular momentum}

\shortauthors{DeFelippis, D., et al.}

\begin{document}

\title{The Impact of Galactic Winds on the Angular Momentum of Disk Galaxies\\in the Illustris Simulation}

\author{Daniel DeFelippis\altaffilmark{1}, Shy Genel\altaffilmark{2,1}, Greg L.~Bryan\altaffilmark{1,2}, S.~Michael Fall\altaffilmark{3}}

\altaffiltext{1}{Department of Astronomy, Columbia University, 550 West 120th Street, New York, NY 10027, USA}
\altaffiltext{2}{Center for Computational Astrophysics, Flatiron Institute, 162 Fifth Avenue, New York, NY 10010, USA}
\altaffiltext{3}{Space Telescope Science Institute, 3700 San Martin Drive, Baltimore, MD 21218, USA}
\email{d.defelippis@columbia.edu}

\begin{abstract}
Observed galactic disks have specific angular momenta similar to expectations for typical dark matter halos in $\Lambda$CDM. Cosmological hydrodynamical simulations have recently reproduced this similarity in large galaxy samples by including strong galactic winds, but the exact mechanism that achieves this is not yet clear. Here we present an analysis of key aspects contributing to this relation: angular momentum selection and evolution of Lagrangian mass elements as they accrete onto dark matter halos, condense into Milky Way-scale galaxies, and join the $z=0$ stellar phase. We contrast this evolution in the Illustris simulation with that in a simulation without galactic winds, where the $z=0$ angular momentum is $\approx0.6\dex$ lower. We find that winds induce differences between these simulations in several ways: increasing angular momentum, preventing angular momentum loss, and causing $z=0$ stars to sample the accretion-time angular momentum distribution of baryons in a biased way. In both simulations, gas loses on average $\approx0.4\dex$ between accreting onto halos and first accreting onto central galaxies. In Illustris, this is followed by $\approx0.2\dex$ gains in the `galactic wind fountain' and no further net evolution past the final accretion onto the galaxy. Without feedback, further losses of $\approx0.2\dex$ occur in the gas phase inside the galaxies. An additional $\approx0.15\dex$ difference arises from feedback preferentially selecting higher angular momentum gas at accretion by expelling gas that is poorly aligned. These and additional effects of similar magnitude are discussed, suggesting a complex origin of the similarity between the specific angular momenta of galactic disks and typical halos.
\end{abstract}

\keywords{galaxies: formation --- structure --- fundamental parameters --- kinematics and dynamics --- methods: numerical --- hydrodynamics}

\section{Introduction}
\label{s:intro}
Understanding the origin of Hubble's tuning fork for galaxy morphological classification is a holy grail of galaxy formation research. It is now known that the morphological classification of a galaxy as an early-type or late-type is strongly correlated with a basic dynamical quantity -- its specific angular momentum content, namely angular momentum per unit stellar mass \citep{FallS_83a,RomanowskyA_12a,ObreschkowD_14a,CorteseL_16a}. This quantity scales with galaxy stellar mass, with two nearly parallel relations existing for late-type galaxies and early-type galaxies, the former having approximately five times as much angular momentum as the latter at a given stellar mass \citep{FallS_13a}. In fact, the angular momentum of a galaxy may well be the more fundamental parameter that is actually driving its morphology. This possibility is receiving increasing attention and scrutiny in recent years thanks to increasingly complete and accurate measurements of galaxy angular momentum content (e.g.,~\citealp{BurkertA_16a,ContiniT_16a,SwinbankM_17a,HarrisonC_17a}), which are much more laborious than those of galaxy morphology. Hence, understanding the origin of galaxy angular momentum will represent a major advance in our understanding of galaxy formation as a whole.

The tight scaling relation between specific angular momentum and stellar mass can be combined with empirical models connecting galaxies to dark matter halos, and with the properties of halos from theory or simulations, to make a statistical connection between the angular momentum contents of galaxies and those of halos. The conclusion from such exercises is that galactic disks have approximately the same values of specific angular momentum as do their host halos (e.g.,~\citealp{ZavalaJ_08a,SokolowskaA_17a}). It is this fact that allows analytical and semi-analytical models that build upon $\Lambda$CDM hierarchical formation to succeed in reproducing various disk galaxy scaling relations provided that they make a simple assumption: that the angular momentum obtained by dark matter halos from cosmological tidal torques is effectively `retained' by the baryons that fall from the circum-galactic medium into the centers of those halos where they form the stellar bodies of galaxies \citep{FallS_80a,MoH_98a}.

This simple assumption, `angular momentum retention', has however historically not been born out in more detailed dynamical models, namely cosmological hydrodynamical simulations. In those simulations, baryons tended to lose the lion's share of the angular momentum they acquired in the intergalactic medium before virial collapse, resulting in unrealistically small galaxies \citep{NavarroJ_95a}. Very recently, however, this situation has changed, with the advent of more accurate solvers \citep{SijackiD_12a}, increased resolution \citep{GovernatoF_04a}, and the introduction of strong galactic winds in the models \citep{SommerLarsenJ_99a,MallerA_02b}. A number of groups have managed to form galactic disks with realistic properties, including size and angular momentum content, in `zoom-in' cosmological simulations (e.g.~\citealp{GrandR_17a}). Moreover, with the increase in computing power, very recent simulations followed large cosmological volumes that contain up to hundreds of massive disk galaxies, and found realistic angular momentum contents not only in a handful of galaxies, but in galaxy populations (e.g.~\citealp{TekluA_15a}). In particular, they are able to reproduce the parallel scaling relations of angular momentum versus stellar mass displayed by observed early-type and late-type galaxies \citep{ZavalaJ_16a}. These advances open the door to detailed studies that will elucidate the nature of angular momentum evolution in a fully (hydro)dynamical cosmological context (e.g.~\citealp{StevensA_17a,LagosC_17a,PenoyreZ_17a}).

In this paper we focus on the high degree of `angular momentum retention' of galactic disks. The starting point for this study is the result of \citet{GenelS_14b} that: i) the population of late-type galaxies in the Illustris simulation has a similar mean angular momentum content to the mean of both their own dark matter halos and observed late-type galaxies, and ii) galactic angular momenta are lower by a factor of a few when galactic winds are turned off. The specific scope of this paper is to describe in what way the galactic winds in the Illustris simulation change the angular momentum evolution of the baryons that make up the stellar components of $z=0$ late-type galaxies at the Milky Way mass scale.

Several ways in which galactic winds may increase the final angular momentum content of a galaxy have been identified in `zoom-in' simulations. First, galactic winds in these simulations preferentially remove gas that has lower specific angular momentum than the mean, hence continuously increasing the mean specific angular momentum of the remaining gas, and consequentially of newly-born stars \citep{GovernatoF_10a,BrookC_10a,OkamotoT_13a,AgertzO_16a}. Second, some fraction of the gas ejected into a galactic wind has been found to fall back to the galaxy (`galactic/halo fountain') with higher angular momentum than that with which it left the galaxy \citep{BrookC_12a, UeblerH_14a,ChristensenC_16a}. Third, in the presence of feedback, galaxies are more gas rich than without feedback and hence baryons lose less angular momentum during galaxy mergers \citep{BrookC_04a,SpringelV_05b,RobertsonB_06a,HopkinsP_09b}. The emerging picture from these works is qualitatively consistent, which is encouraging given that they were based on different hydrodynamics codes, feedback schemes, and mass scales. However, these differences, as well as the small number of isolated galaxies included in these analyses, imply that no comprehensive, detailed, and quantitatively consistent picture exists as of yet.

The main focus of the present work is a quantification of the changes that the angular momenta of baryons comprising the stars in late-type galaxies undergo between the time they entered their host halos and $z=0$. We define several distinct `events' in the evolution of every baryonic mass element and divide this full time period into several intervals using these events. We then compare the angular momentum evolution during those intervals between the Illustris simulation and a similar simulation run without galactic winds. This study focuses on providing answers to `When', `Where', and `How much', setting the stage for future studies of the `How' and `Why'.

With respect to the existing literature on this topic, the tools used in this work are unique in two aspects. First, it is based on a large population of simulated galaxies in a simulation that reproduces observed angular momentum relations \citep{GenelS_14b} as well as many other properties of galaxy populations \citep{VogelsbergerM_14a,GenelS_14a,TorreyP_14a}. Second, it employs a Lagrangian analysis in a simulation based on a mesh code, using tracer particles, while previous Lagrangian analyses on this topic have all been based on Smoothed Particle Hydrodynamics (SPH) (e.g.~\citealp{ZavalaJ_16a}).

This paper is organized as follows. In Section \ref{s:methods} we describe the simulations and our analysis methodologies. In Section \ref{s:results} we present the evolution of angular momentum and contrast the two types of simulations, with and without galactic winds. Section \ref{s:results_breakup} is the main results section and \Fig{summary} presents its key plot. In Section \ref{s:summary} we discuss our results within a broader context and summarize them.

\section{Methods}
\label{s:methods}
\subsection{Simulations}
\label{s:simulations}
We use the Illustris-2 simulation \citep{GenelS_14a,VogelsbergerM_14a,VogelsbergerM_14b} of a $(106.5\Mpc)^3$ volume, as well as a `No-Feedback' simulation of a $(35.5\Mpc)^3$ volume, both evolved with a WMAP-9 $\Lambda$CDM cosmology \citep{HinshawG_13a} down to $z=0$ using the moving-mesh code \AREPO{ }\citep{SpringelV_10a}. The former is initialized with $910^3$ dark matter and baryonic resolution elements, and the latter with $256^3$, implying that they have similar resolutions in space ($\sim\kpc$) and mass ($\sim(1-2)\times10^7\Msun$ baryonic; $\sim(5-10)\times10^7\Msun$ for dark matter). Both simulations include gas cooling and stochastic star formation, but only Illustris-2 (hereafter `Illustris') has feedback in the form of star formation-driven galactic winds, as well as black hole formation and evolution \citep{VogelsbergerM_13a}, rendering No-Feedback very similar to the simulations in \citet{VogelsbergerM_12a}. For the purposes of this work, and in particular with regards to angular momentum, the results at this resolution level are converged well enough with respect to the higher-resolution Illustris-1 simulation and its no-feedback analogue (for detailed resolution studies see \citealp{VogelsbergerM_12a,VogelsbergerM_13a,GenelS_14b}).

The implementation of galactic winds in Illustris closely follow the technique introduced in \citet{SpringelV_03a}. Wind particles are launched stochastically directly from the star-forming gas with prescribed velocities and mass-loading factors that depend on the local dark matter velocity dispersion around the star-forming cells, which itself closely follows the local gravitational potential. The wind ejection velocities are set to be larger than the escape velocity from the galaxy but typically smaller than the escape velocity from the host halo, such that wind particles typically reach maximum distances that are comparable to but smaller than the virial radii of their host halos. The mass-loading factors are derived from the wind velocity such that the kinetic energy associated with the ejections per unit star-formation rate is a constant that corresponds to $\approx3\times10^{51}\erg$ per supernova. This results in mass-loading factors that are typically greater than unity, and on the order of $5$ for galaxies around the Milky Way mass, as is commonly employed in cosmological simulations with comparable resolution to Illustris \citep{ZahidJ_13b}\footnote{These mass-loading factors are high compared to direct observational estimates, however they should not be compared at face value. Beyond the large uncertainties on observational mass-loading measurements, they are measured at a distance from the disk while the simulated mass-loading factors apply directly at the ejection from the disk. A robust comparison of mass-loading factors between simulations and observations is beyond the scope of this work.}. The wind particles are first decoupled from hydrodynamical forces and move ballistically to allow them to escape the galaxies, and are recoupled to the gas after either a short amount of time or when they reach a low density region. The direction of the momentum kick given to a wind particle is perpendicular to both the velocity and the acceleration of the star-forming cell from which it is launched with respect to the galaxy center. All these various aspects of the implementation and the numerical values of the adjustable parameters were set with the aim of approximately reproducing the stellar mass function of galaxies at $z=0$ and the global history of cosmic star-formation density. No aspect of the angular momentum of galaxies was tuned for.

Halos are found with the friends-of-friends algorithm (\FOF, \citealp{DavisM_85a}). \FOF{ }halos may have general shapes and their boundaries roughly trace a constant density contour such that their mean density corresponds roughly to $200$ times the mean cosmic matter density (for relations between halo definition and angular momentum, see e.g.~\citealp{ZjupaJ_16a}). Galaxies are identified using the \SUBFIND{ }algorithm \citep{SpringelV_01}. These are gravitationally bound objects constructed around density peaks. We define a `galaxy' as the collection of all stellar particles, as well as gas particles with a density above $0.13\pcmcb$ -- referred to as `star-forming gas', inside any given \SUBFIND{ }object. The data from each simulation include $136$ snapshots and corresponding group catalogs, providing a time resolution of the order of $100\Myr$.

We utilize a Lagrangian point of view for the evolution of angular momentum, meaning that we are interested in the angular momentum histories of unique baryonic mass elements as they travel across cosmic time from the uniform initial conditions through the cosmic web into dark matter halos and finally into the galaxies where they reside at the present epoch. To perform a Lagrangian analysis in a mesh-based code like \AREPO{ }requires using tracer particles, since the hydrodynamical cells represent a discretization of space, not of mass. These are implemented using the Monte Carlo method introduced in \citet{GenelS_13a}, where each tracer belongs at any given time to a certain baryonic resolution element (including gas cells, stellar and black hole particles, as well as wind particles). These `passive' tracers carry only their identity throughout the simulation, and no mass, however they do continuously record certain properties of the cells they belong to. For example, a property that we use in this work is the `wind-counter' each tracer stores, which increases by unity every time that tracer is incorporated into the galactic wind. Following the Lagrangian evolution of the angular momentum of certain $z=0$ galaxies means following back in time the tracers associated with the `active' baryonic elements comprising those galaxies.

\subsection{Analysis}
\label{s:analysis}
\begin{figure*}
\centering
          \includegraphics[width=1.0\textwidth]{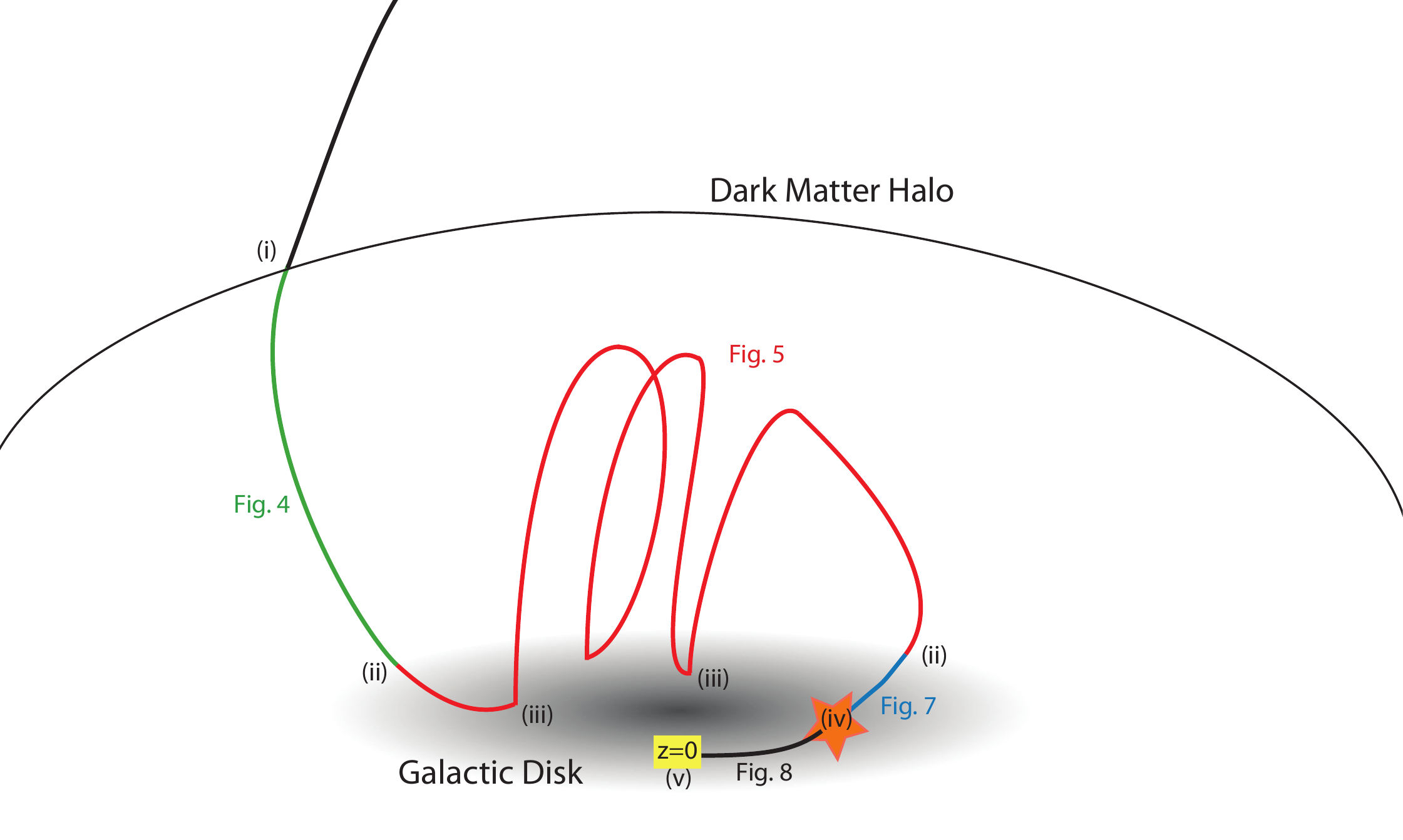}
\caption{A cartoon illustrating the various `events' in the evolution of a tracer particle that are considered in this work. These are (i) halo accretion, (ii) first/last becoming part of the star-forming phase, (iii) first/last ejection into the wind, (iv) becoming part of the stellar phase, (v) $z=0$. In addition, certain intervals between these events that are addressed by particular figures are marked as such.}
\vspace{0.3cm}
\label{f:cartoon}
\end{figure*}

The main analysis tool we present in the next section is relationships between the angular momentum values of individual tracers at particular `events' in their evolution history. Generally, for each tracer each of these events occurs at a different cosmic time. We make direct comparisons of identical event types between Illustris and No-Feedback, and also examine certain types of events that only occur in Illustris, namely those related to the galactic winds. The events are illustrated in \Fig{cartoon} and defined as follows:
\renewcommand{\labelenumi}{(\roman{enumi})}
\begin{enumerate}
\item Accretion onto main halo: the snapshot when a tracer first becomes part of the FOF{ }halo that is on the main progenitor branch of the FOF{ }halo it ends up in at $z=0$.
\item First (last) star-forming gas: the snapshot when a tracer is first (last) recorded {\it entering} the star-forming gas phase, namely crossing from below a density threshold of $0.13\pcmcb$.
\item First (last) ejection: the snapshot when a tracer is first (last) recorded switching from a gas cell to a wind particle (only defined for Illustris).
\item Star-formation: the snapshot when a tracer is last recorded changing from a gas cell or wind particle to a stellar particle.
\item $z=0$ star: the final snapshot in the simulation (defined only for tracers that belong to the stellar component at that time). 
\end{enumerate}

All tracers that belong to a stellar particle at $z=0$ are included in the analysis (which focuses our analysis on the stellar angular momentum), except those that join the central galaxy while already belonging to a star particle. This latter criterion excludes `ex-situ' stars, which are not directly affected by the winds and are therefore left outside the scope of this paper. This removes 10\% of the $z=0$ stars in Illustris and 40\% in No-Feedback. For those tracers that are included in the analysis, we exclude events -- except accretion onto the halo -- that occur while a tracer is contained in a satellite galaxy. This is because as the angular momentum of these tracers with respect to the main progenitor galaxy is dominated by the orbital angular momentum of the satellite and hence not meaningful for our purposes. In other words, the starting point for the events above in the time line of each tracer is the time it becomes part of the main progenitor halo. The `accretion onto main halo' event is the exception, as it is considered also for tracers that accrete as part of a satellite.

To calculate the specific angular momentum of a particle, we define a center of rotation as the minimum of the potential well of its host galaxy \citep{GenelS_14b}. To do this at all simulation snapshots, we use the \SUBLINK{ }merger trees \citep{Rodriguez-GomezV_14a} to find the main progenitor branch of the $z=0$ galaxy and calculate the angular momentum of all particles with respect to the main progenitor, regardless of whether the particles already belong to that main progenitor or are yet to be accreted onto it. We also define the reference frame for the angular momentum calculation as having the velocity of the center of mass of the main progenitor, and specifically of all the stars and star-forming gas present in the main progenitor galaxy. Then, we calculate the specific angular momentum of a tracer particle $i$ as follows:
\be
    \mathbf{j}_{i} = (\mathbf{r}_{i} - \mathbf{r}_{\rm{minpot}}) 
    \times (\mathbf{v}_{i} - \mathbf{v}_{\rm{COM}}).
\label{e:j_tracer}
\ee
To compute the total angular momentum of a galaxy we sum the angular momenta of the tracers associated with it\footnote{This is a simple rather than a weighted average because all tracer particles represent equal masses. It is not an exact value due to the Monte Carlo noise in the number of tracers per cell (see \citealp{GenelS_13a}), but for our galaxies of interest, with tens of thousands of resolution elements, this is an excellent approximation.}:
\be
    \mathbf{j}_{\mathrm{gal}} = \frac{1}{N_\mathrm{tr,gal}} \sum\limits_{i=1}^{N_{\mathrm{tr,gal}}} \mathbf{j}_i.
\label{e:j_galaxy}
\ee
As the angular momentum is a (pseudo-)vector, the magnitude of the sum and the sum of the magnitudes are different quantities, when looking at many tracers together. In the next section, we find both quantities to be informative, as well as the comparison between them. We define the level of self-alignment of a population of tracers as the ratio of these quantities,
\be
    A=\frac{|\sum\mathbf{j}_i|}{\sum|\mathbf{j}_i|}.
\label{e:j_alignment}
\ee
If all vectors of a particular tracer population cancel each other out, the self-alignment is $A=0$, while if they all have the same direction, the self-alignment equals $A=1$. 

When calculating the vector sum across different galaxies, one has to take into account the fact that each galaxy is in general oriented in a different direction in the simulation box, such that simply summing different galaxies together will necessarily lead to a meaningless vector cancellation. Hence, for the purpose of summing up individual tracer vectors across many galaxies, each $\mathbf{j}_{i}$ at any particular event is measured as in \equ{j_tracer} but in a reference frame that is rotated such that the $z$ axis points in the direction of $\mathbf{j}_{\mathrm{gal}}$ of the galaxy hosting the tracer at the time of that event.

In order to focus the scope of the paper, we select central galaxies at the Milky Way mass scale, namely with virial masses \citep{BryanG_97a} in the ranges $10^{12.1}<M_h[\Msun]<10^{12.2}$ and $10^{11.65}<M_h[\Msun]<10^{12.65}$ for Illustris and No-Feedback respectively. The mass range used for No-Feedback is larger, given its smaller cosmic volume, as these bins are chosen to follow an equal total number of tracers in each simulation, $\approx10^6$. This selection results in $278$ galaxies in Illustris and $140$ galaxies in No-Feedback. In order to further narrow our focus to disk galaxies, we consider only the galaxies that are at the high-tail of the angular momentum distribution, which correspond to disks both from observations \citep{RomanowskyA_12a} and in Illustris \citep{GenelS_14b}. Specifically, we select the 25\% of central galaxies with the highest stellar angular momentum at $z=0$ in both simulations.

\section{Results}
\label{s:results}
\subsection{The overall picture from accretion onto halos to $z=0$}
\label{s:results_overall}
\begin{figure*}
\centering
\subfigure[]{
          \label{f:Ill2-z0acc}
          \includegraphics[width=0.314\textwidth]{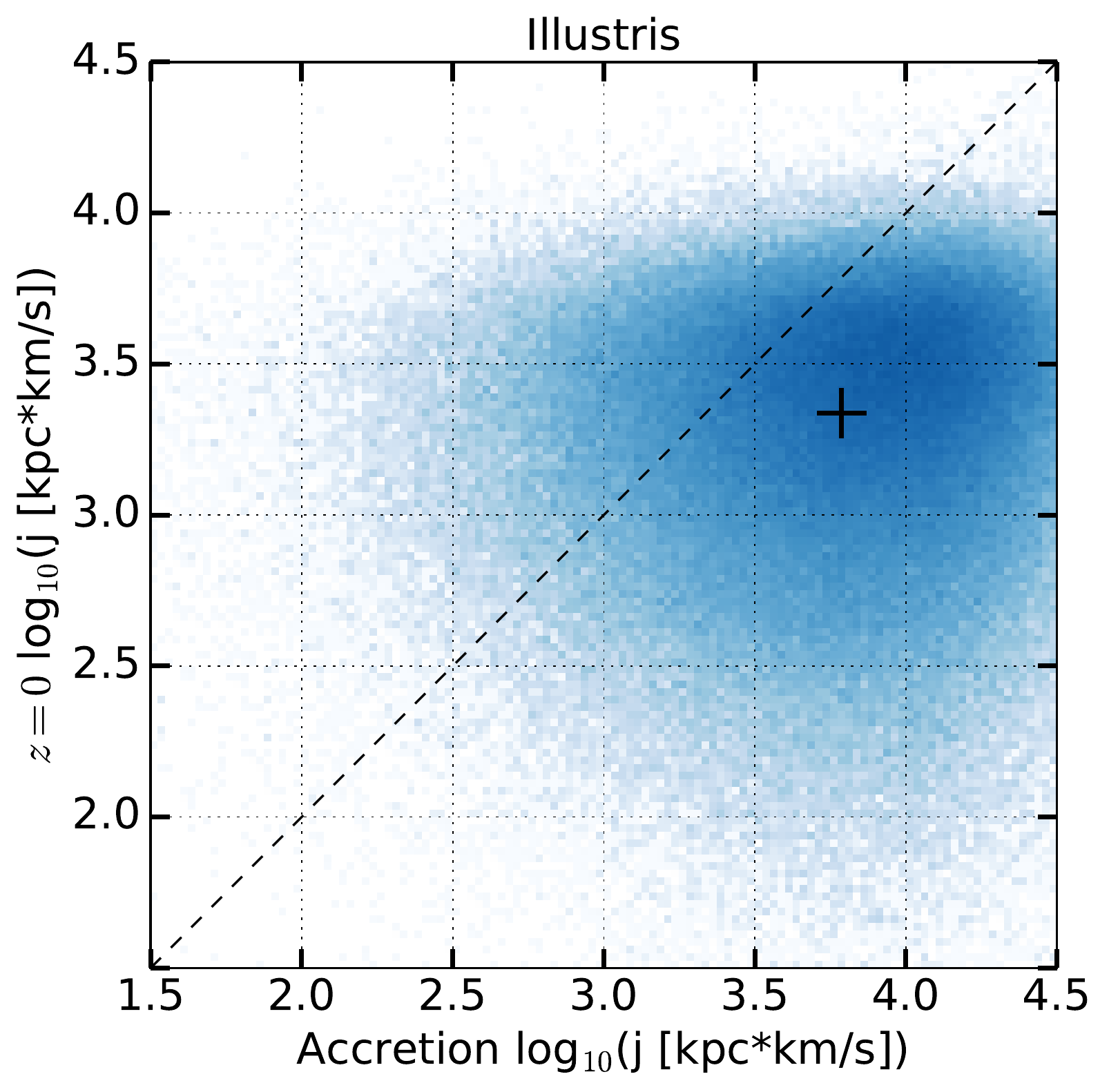}}
\subfigure[]{
          \label{f:NoFeed-z0acc}
          \includegraphics[width=0.3215\textwidth]{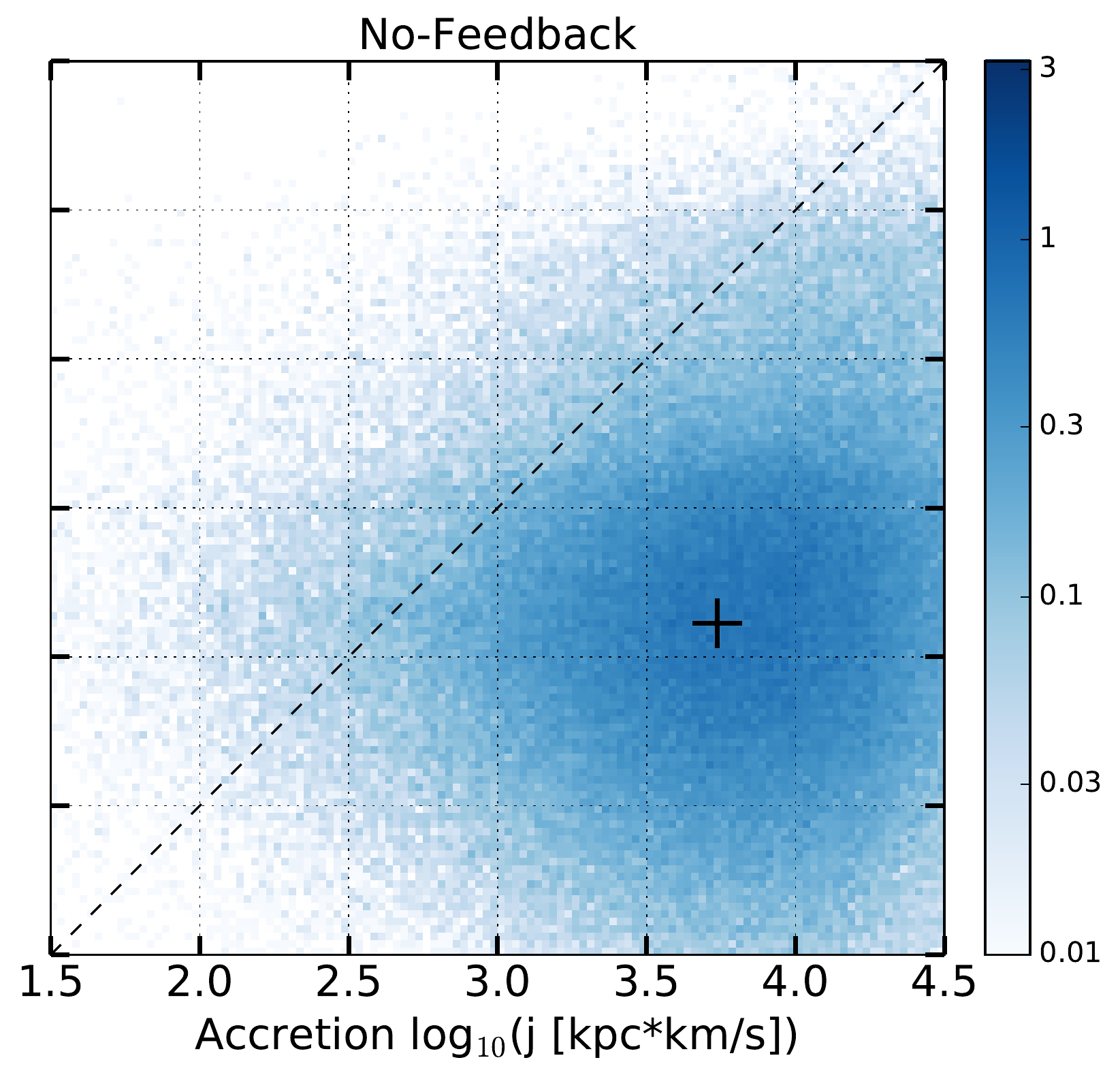}}
\subfigure[]{
	\label{f:z0acc1D}
	\includegraphics[width=0.3355\textwidth]{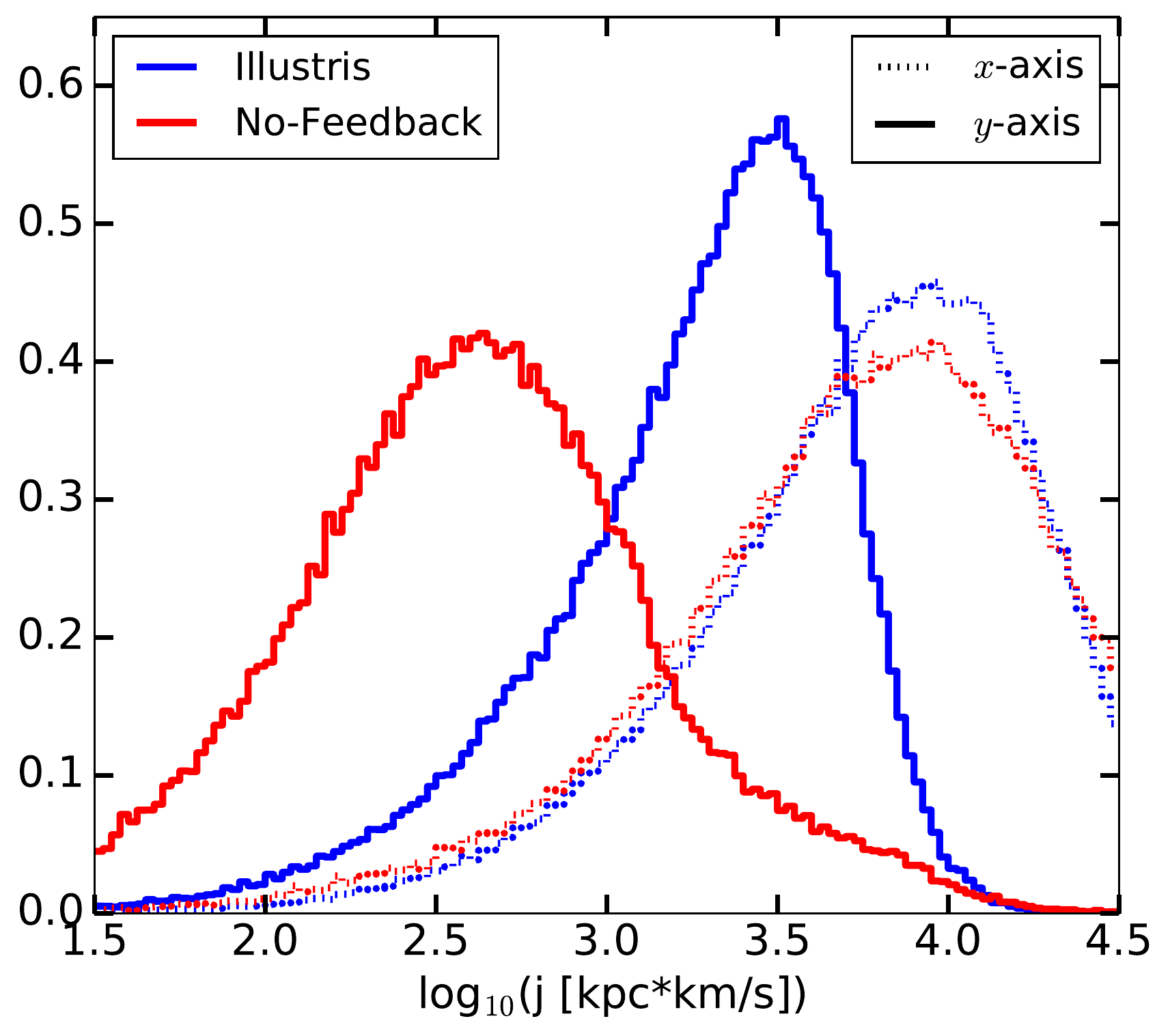}}
\caption{Joint (left and middle) and one-dimensional  (right) probability distributions of angular momentum magnitudes of stellar tracers at $z=0$ (vertical axes) and those same tracers as gas at their time of accretion onto the host halo (horizontal axes). The units indicated by the color bars are of probability per $\dex^2$. The diagonal dashed lines indicate the $1:1$ relation. The cross indicates the median angular momentum at both events. In both simulations, an overall average loss of angular momentum is evident, but to a much larger degree in the simulation without feedback ($0.65\dex$; middle) than in Illustris ($0.23\dex$; left).} 
\vspace{0.3cm}
\label{f:z0acc}
\end{figure*}

In \Fig{z0acc} we present the joint, two-dimensional distributions of the magnitudes of the specific angular momentum vectors of individual tracers at two distinct events, both for Illustris (left) and No-Feedback (middle). For each simulation, the tracers included in these distributions are all those that are part of the stellar component of $z=0$ galaxies selected as described in Section \ref{s:methods}. The vertical axes represent the angular momentum of each tracer at a fixed cosmological time, $z=0$. The horizontal axes represent the angular momentum at the time of accretion onto the halo (which occurs in general at different times for different tracers). The striking difference between the two simulations on the vertical axis is in essence the result of \citet{GenelS_14b} that in a simulation without feedback, galaxies at $z=0$ have $\approx0.5\dex$ lower angular momentum content. This is the result that motivates this work. The difference between the two simulations on the horizontal axis is much milder. This suggests that most of the difference represented on the vertical axis develops in between these two events, namely inside halos, rather than before the accretion onto them. In Section \ref{s:results_breakup} and the following figures (as indicated in \Fig{cartoon}), we will break this typically long time interval between accretion onto the halo and $z=0$ into sub-intervals and examine each of them separately, which will constitute the main results. The difference on the horizontal axis that represents the earliest event considered in our main analysis is not zero, and we will return to it in Section \ref{s:results_bias}, but it is mild, representing a similar starting point to the main analysis between the two simulations.

Examining each panel in \Fig{z0acc} by itself, it is worth noting that in Illustris the magnitudes of specific angular momentum loss (around the peak of the distribution, where most tracers are located) between accretion and $z=0$ range between $\approx0-1\dex$, and in No-Feedback the corresponding losses are $\approx0.5-2\dex$. Put more precisely, we find that the sum of the angular momentum magnitudes at $z=0$ is lower than the sum of angular momentum magnitudes at halo accretion time by $0.57\dex$ in Illustris and by as much as $1.04\dex$ in No-Feedback. Additionally, in both panels, the peak of the distribution occupies a locus that is significantly shallower than the $1:1$ relation. This means that gas accreted onto the halo with high angular momentum tends to lose by $z=0$ a larger fraction of that angular momentum compared with gas that was accreted with lower angular momentum in the first place. 

The sum of vector magnitudes does not tell the full story, however. If instead we examine the magnitudes of the vector sums on each of the axes in \Fig{z0acc}, we find that the difference between $z=0$ and halo accretion time is only $0.23\dex$ in Illustris and $0.65\dex$ in No-Feedback. That these numbers are smaller than the differences of the magnitude sums ($0.57\dex$ and $1.04\dex$ respectively, as reported in the previous paragraph) means that the individual vectors are significantly more aligned at $z=0$ than at the halo accretion time. This by itself is easy to understand as a result of angular momentum cancellation. The angular momentum magnitudes of {\it individual} tracers drop by a combination of: {\bf(i)} transport of angular momentum to other, potentially both baryonic and dark matter, components (which accounts for the decrease of the magnitude of the vector sum), and {\bf(ii)} cancellation with other baryons that end up in the $z=0$ galaxy but have been accreted with different angular momentum directions (which does not change the magnitude of the vector sum).

To summarize, in terms of the magnitude of the total specific angular momentum vector, baryons experience a significant angular momentum loss between the time when they are accreted onto halos and $z=0$ in the no-feedback simulation ($0.65\dex$), an effective loss that is much smaller in the Illustris simulation ($0.23\dex$). In the following sub-section we break this difference to smaller intervals in order to gain insight into its nature and origin.

\subsection{The evolution between various events}
\label{s:results_breakup}
\begin{figure}
\centering
           \includegraphics[width=0.49\textwidth]{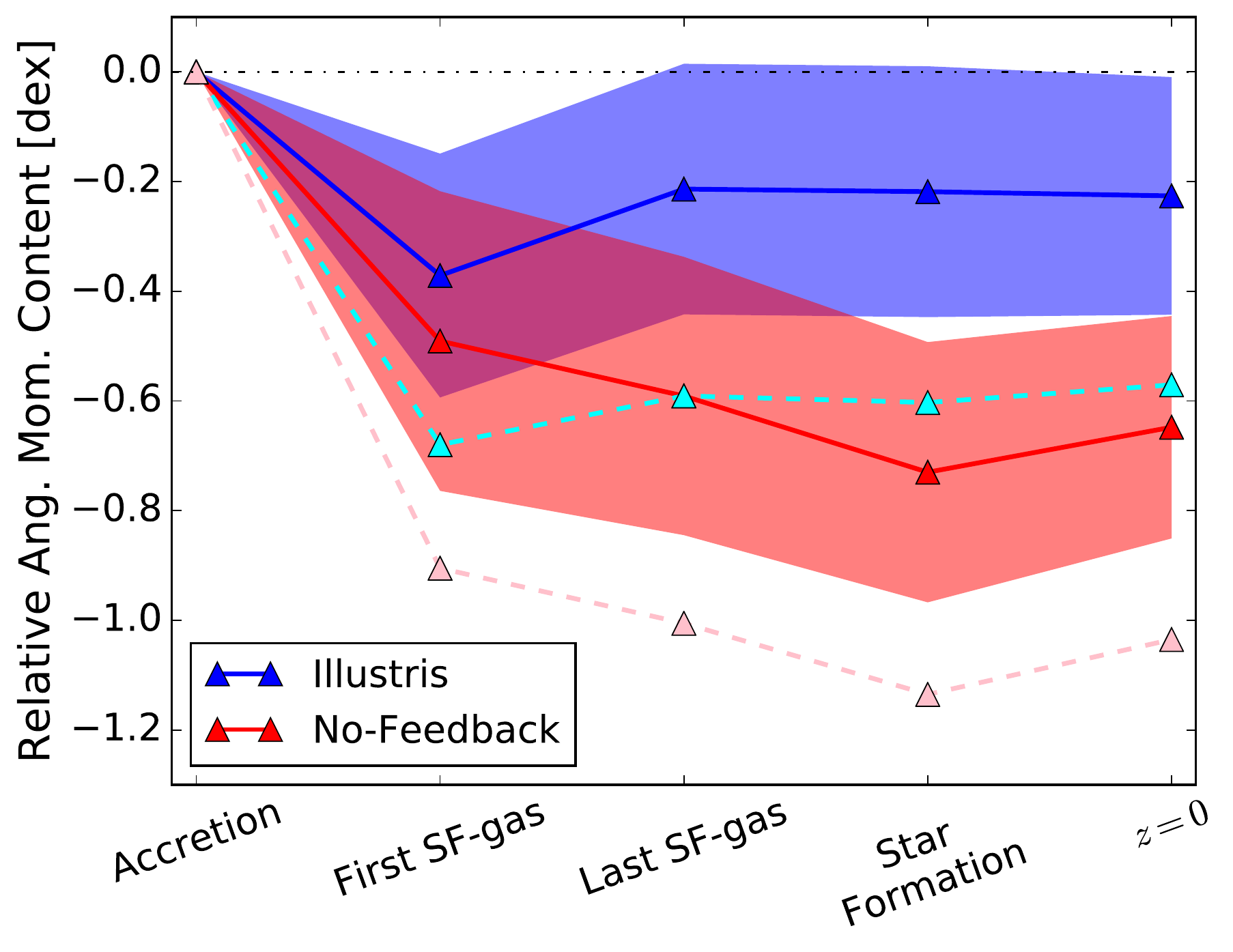}
\caption{The average angular momentum loss of tracers at events defined in Sec.~\ref{s:analysis} relative to their accretion value for Illustris (blue/cyan) and No-Feedback (red/pink). Solid lines in dark color show the magnitude of the vector sum, while dashed lines in light color show the mean magnitude. The dark-shaded regions are the $1\sigma$ spread of relative losses among different galaxies. The $1\sigma$ spreads of the relative losses of each individual tracer are $\sim1\dex$ and are not shown for clarity.} 
\vspace{0.3cm}
\label{f:summary}
\end{figure}

\Fig{summary} shows a quantitative summary of the results presented in this section for the convenience of the reader. The horizontal axis represents the sequence of events defined in Section \ref{s:analysis} and \Fig{cartoon}, with the events discussed in Section \ref{s:results_overall} and \Fig{z0acc} shown as the initial and final points. The vertical axis represents the difference in angular momenta, in logarithmic space, relative to the starting point of accretion onto the halo. The nearly-monotonic loss of mean angular momentum in No-Feedback (red) and non-monotonic evolution in Illustris (blue) are shown in more detail in the following Figs.~\ref{f:firstSFacc} through \ref{f:redshift0laststar}.

In \Fig{firstSFacc} the horizontal axes show the same quantity as the horizontal axes in \Fig{z0acc}, namely the angular momentum at the time of halo accretion. The vertical axes show the angular momentum magnitude of each tracer at the time it first crosses the density threshold for star-formation, i.e.~when it first becomes part of the star-forming phase, inside the main progenitor galaxy. This time interval represents the first `halo crossing' from the outskirts to the central part of the halo, and is marked in green in \Fig{cartoon}. The difference between the two simulations is again significant. In Illustris, there is a clear correlation between the angular momentum at the two events. From \Fig{firstSFacc} we read off an approximately constant degree of loss of $\approx(0.5\pm0.2)\dex$, which amounts to an overall loss during this first passage through the halo of $0.68\dex$ in the magnitude sum and $0.37\dex$ in the magnitude of the vector sum (again indicating that some angular momentum cancellation occurs between the virial radius and the galaxy itself). In No-Feedback, on the other hand, the relation is much shallower than the $1:1$ relation, amounting to an overall loss of $0.91\dex$ in the magnitude sum and $0.49\dex$ in the magnitude of the vector sum.

\begin{figure*}
\centering
\subfigure[]{
          \label{f:Ill2-firstSFacc}          
          \includegraphics[width=0.314\textwidth]{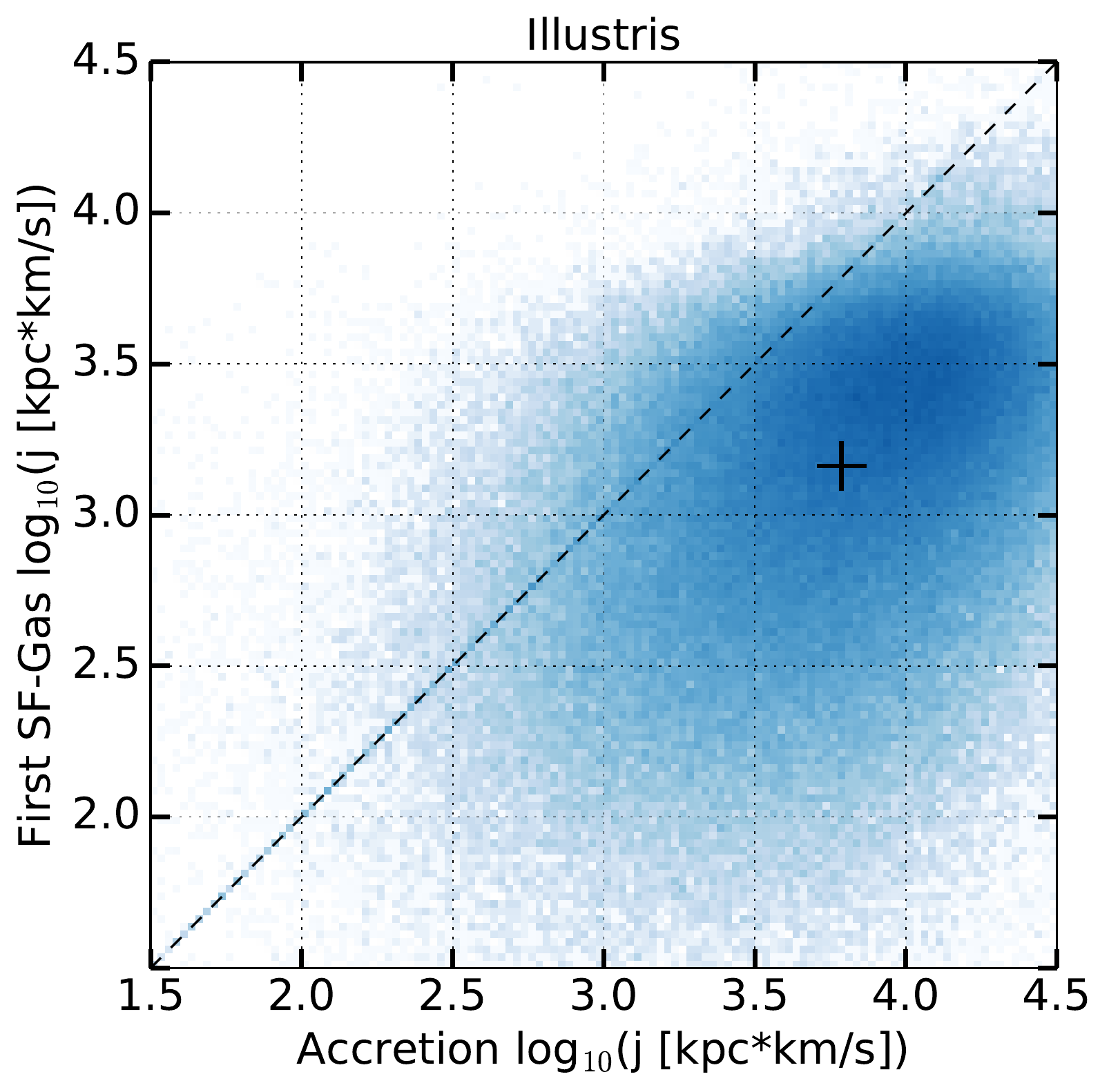}}
\subfigure[]{
          \label{f:NoFeed-firstSFacc}
          \includegraphics[width=0.3215\textwidth]{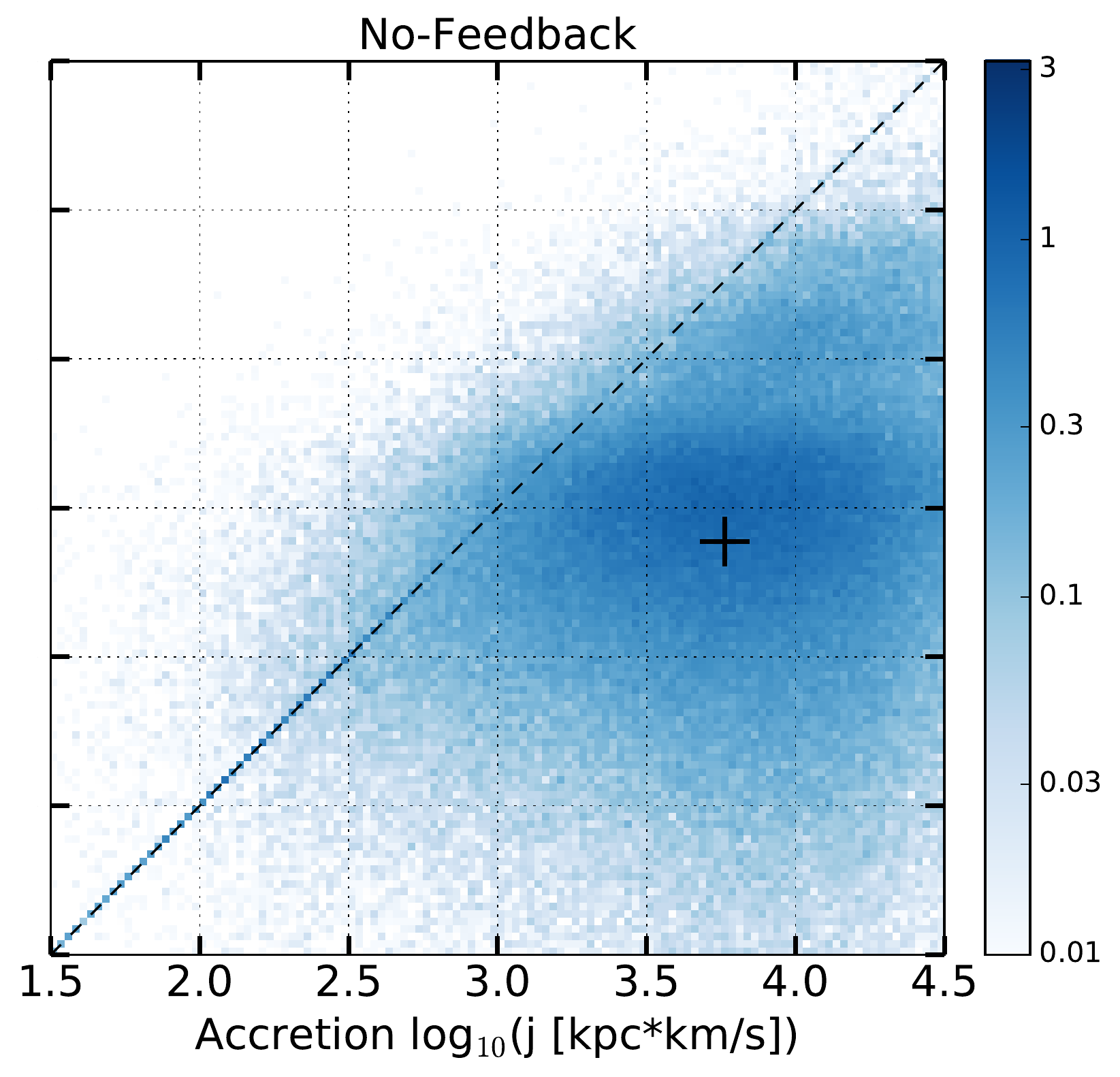}}
\subfigure[]{
          \label{f:firstSFacc1D}
          \includegraphics[width=0.3355\textwidth]{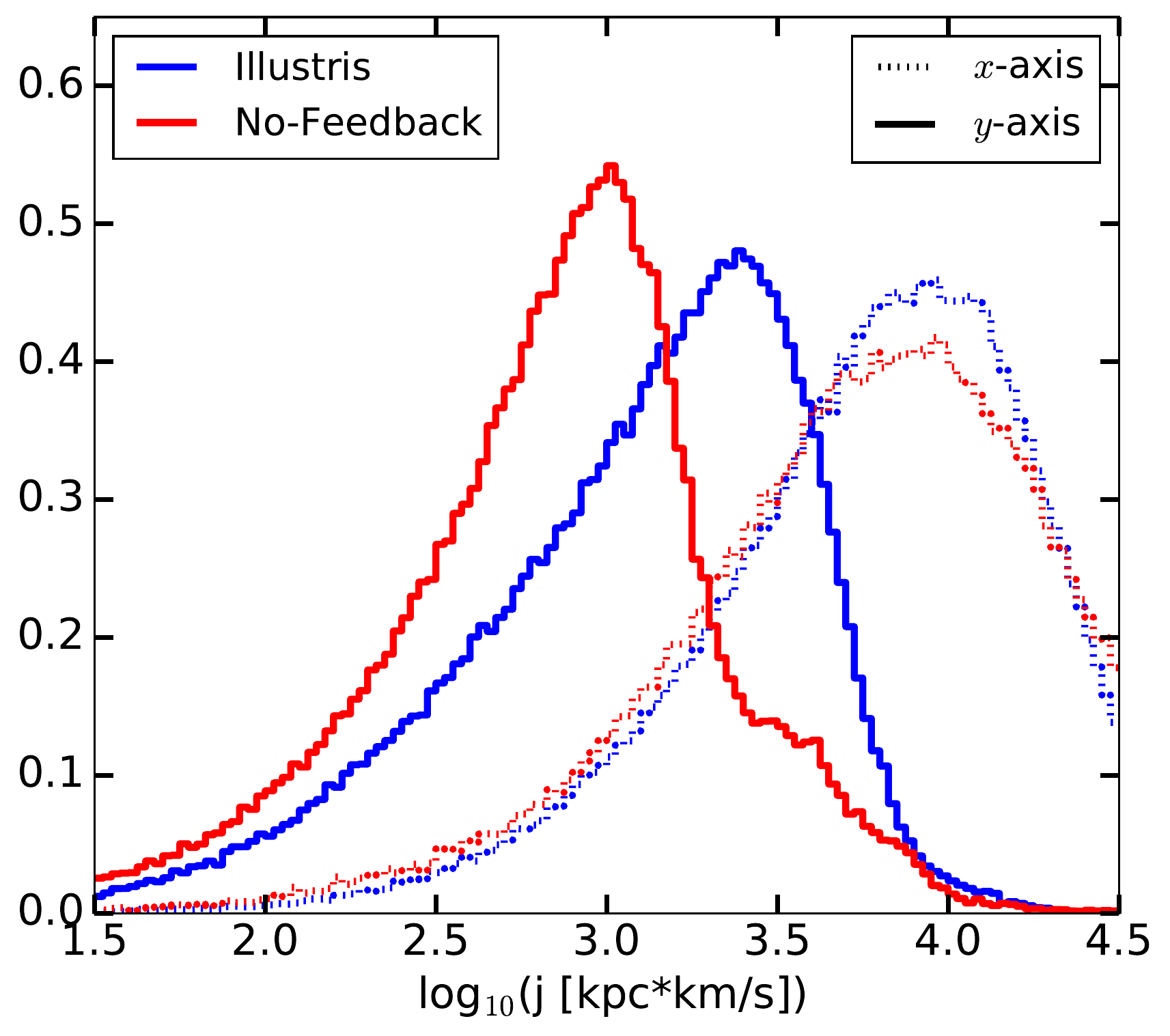}}
\caption{Joint and one-dimensional probability distributions of angular momentum magnitudes of tracers as they first cross the star-formation density threshold (vertical axes) and at accretion onto the host halo (horizontal axes). The No-Feedback plot (middle) resembles its counterpart in \Fig{z0acc} but translated up $\sim 0.4\dex$ on the vertical axis, meaning that only a fraction of the loss seen in \Fig{z0acc} occurs before crossing the star-formation density threshold. The positive slope in the Illustris plot (left), when compared to the flatter one in \Fig{z0acc}, indicates that tracers accreted with $\log(j) \sim 3.5$ lose some angular momentum before crossing the star-formation density threshold and then re-gain it by $z=0$, while tracers accreted with $\log(j) \sim 4$ already have here their $z=0$ value.}
\vspace{0.3cm}
\label{f:firstSFacc}
\end{figure*}

That the difference between these two sum measures is larger in No-Feedback ($0.91-0.49=0.42\dex$) than in Illustris ($0.68-0.37=0.31\dex$) means that there is more angular momentum cancellation in No-Feedback. We find that this originates primarily in a lower level of self-alignment (see \equ{j_alignment}) in No-Feedback at the time of halo accretion, $A_{\rm No-FB}^{\rm acc}=0.24$ compared with $A_{\rm Illustris}^{\rm acc}=0.36$, rather than in self-alignment differences at the time of crossing the star-formation threshold, $A_{\rm No-FB}^{\rm SF}=0.62$ and $A_{\rm Ill}^{\rm SF}=0.73$. One might hypothesize that a lower level of self-alignment is a result of a wider distribution of accretion times in No-Feedback; however, the distribution of accretion times is similar between the two simulations (\Fig{redshiftdist}), implying a different origin. This is discussed further in Secion \ref{s:results_bias}.

To summarize the time interval shown in \Fig{firstSFacc}, in terms of the magnitude of the total specific angular momentum vector, the loss is rather large in both simulations and not dissimilar, namely $0.37\dex$ in Illustris and $0.49\dex$ in No-Feedback. Comparing these numbers to those quoted in Section \ref{s:results_overall} based on \Fig{z0acc} ($0.23\dex$ and $0.65\dex$ respectively), we conclude that losses during the first passage through the halo represent roughly three-quarters of the total loss experienced in No-Feedback by $z=0$. In contrast, in Illustris we expect to find a time interval that occurs {\it after} the first crossing of the star-formation threshold during which almost half of the losses incurred before that crossing are counteracted. This is indeed shown in the next two figures.

\Fig{lastSFfirstSF} presents the joint angular momentum magnitude distribution at the last time, versus the first time, tracers join the star-forming phase in the main progenitor of their $z=0$ galaxy, and is marked in red in \Fig{cartoon}. We begin with a discussion of Illustris, where this includes the full time a tracer is in the galactic fountain during which it typically goes out of the galaxy, and falls back in, several times (only $\approx20\%$ of the tracers join the star-forming phase only once). \Fig{lastSFfirstSF} shows that the angular momentum magnitude at the end of this cycle is typically higher than at its beginning, more so for tracers that have low angular momentum magnitudes at the first time they join the star-forming phase. Over the whole tracer population, the mean magnitude increase is $0.09\dex$, and the magnitude of the vector sum increases by $0.16\dex$. This latter gain undoes almost half of the loss that is incurred between halo accretion and arrival at the galaxy (\Fig{Ill2-firstSFacc}), and its origin will be discussed further in relation to \Fig{ejectionreturn}.

\begin{figure*}
\centering
\subfigure[]{
          \label{f:Ill2-lastSFfirstSF}
          \includegraphics[width=0.314\textwidth]{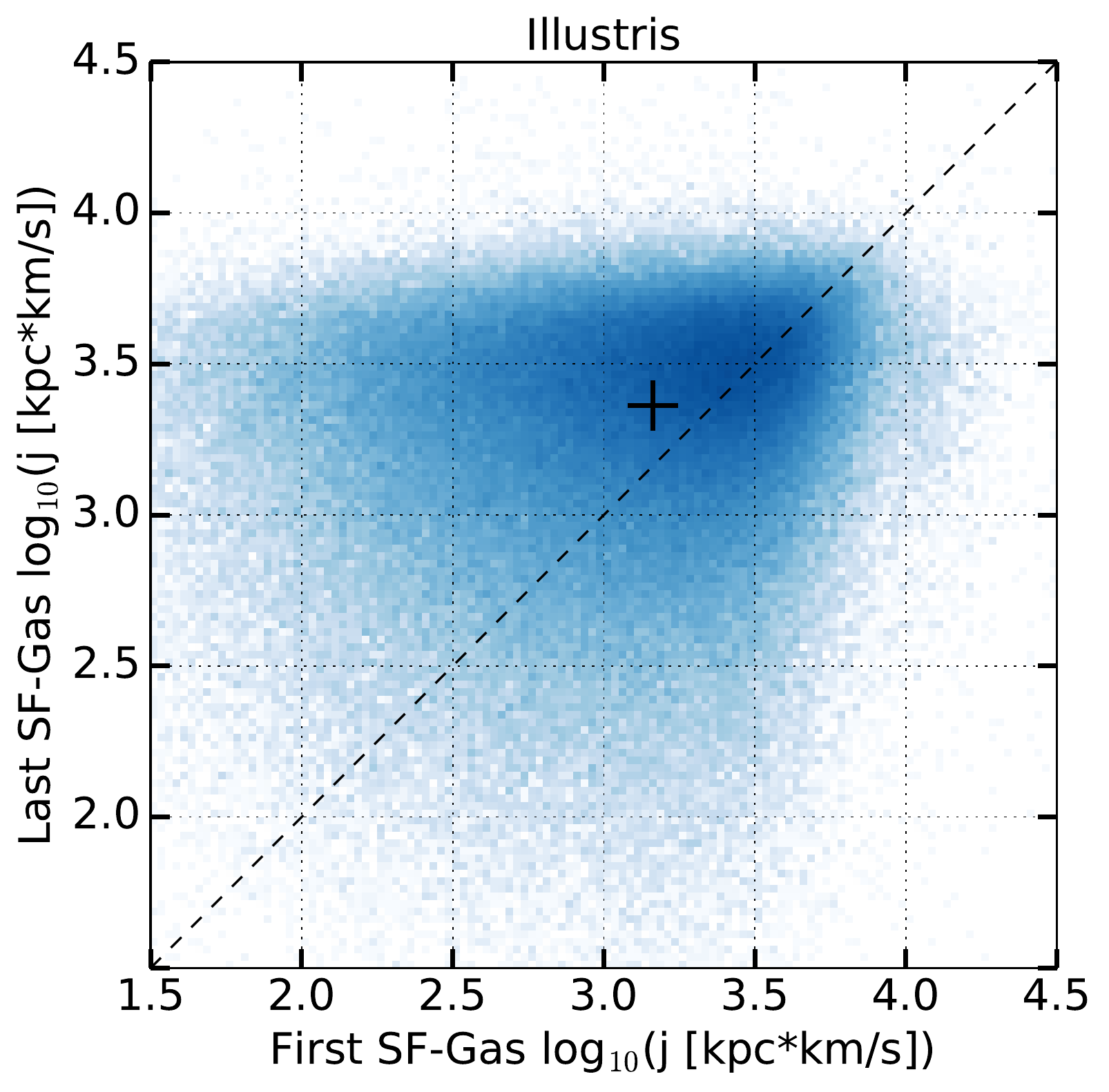}}
\subfigure[]{
          \label{f:NoFeed-lastSFfirstSF}
          \includegraphics[width=0.3215\textwidth]{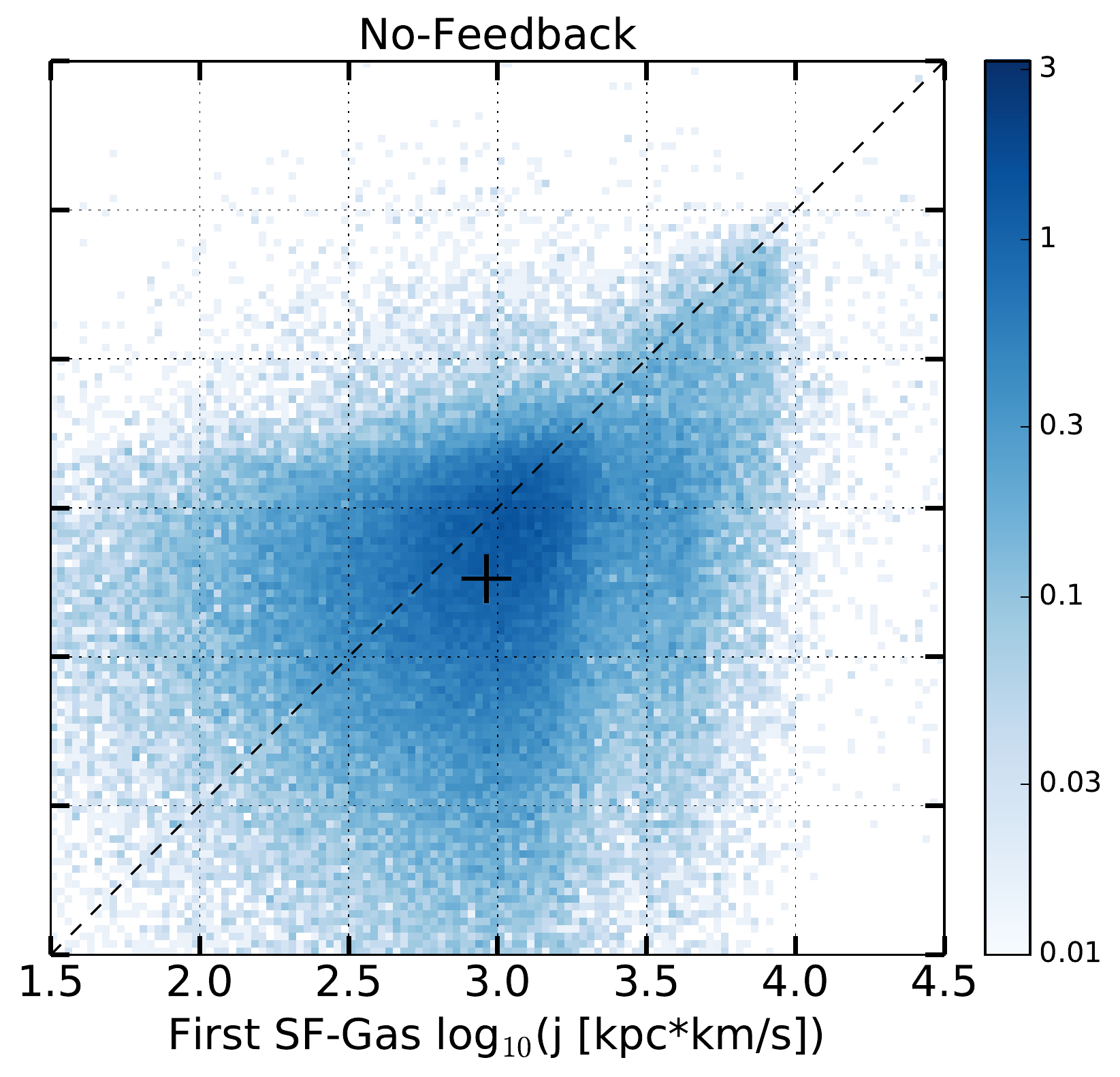}}
\subfigure[]{
          \label{f:lastSFfirstSF1D}
          \includegraphics[width=0.3355\textwidth]{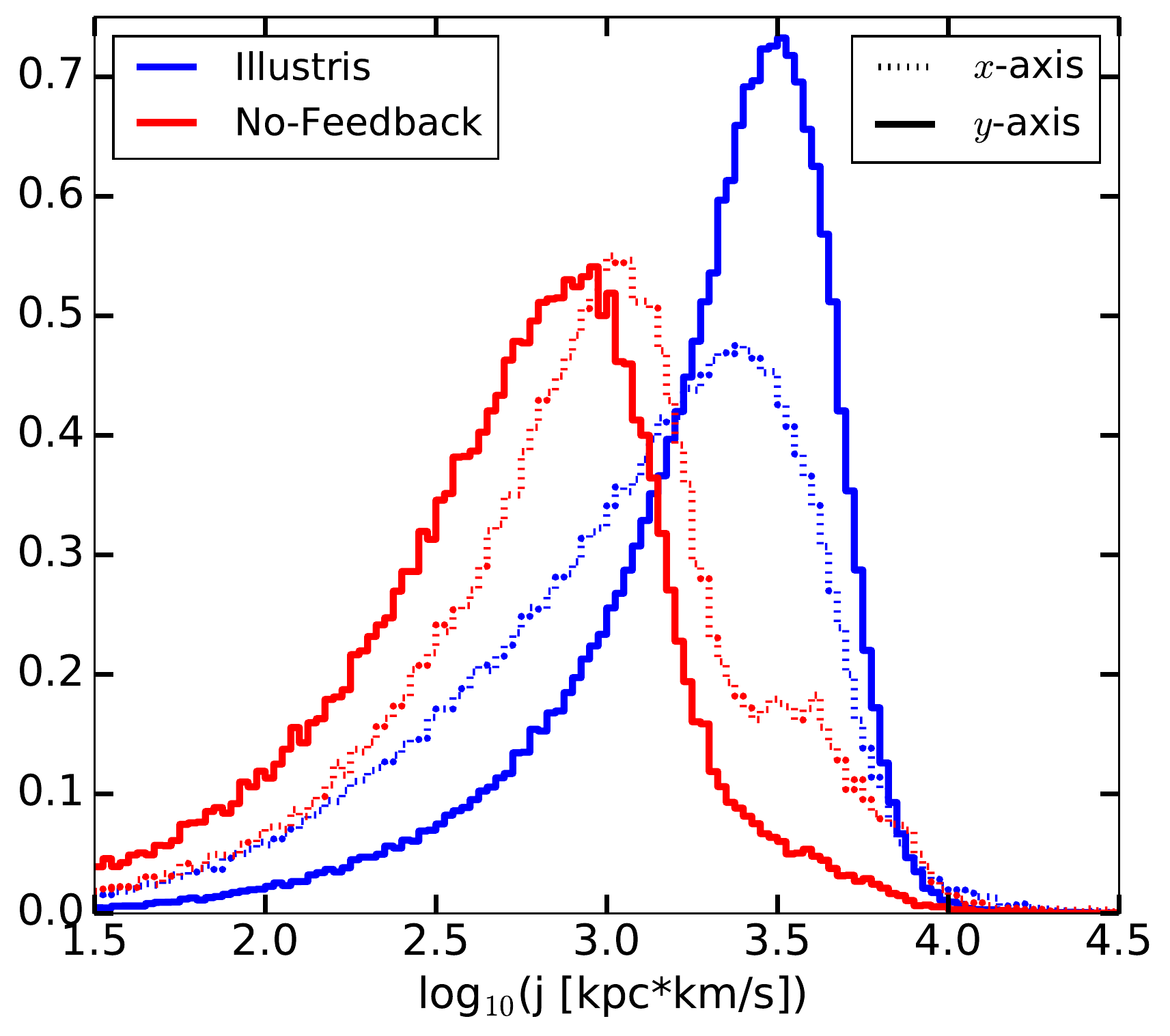}}
\caption{Joint and one-dimensional probability distributions of angular momentum magnitudes of tracers as they first cross the star-formation density threshold (horizontal axes; the same quantity as on the vertical axes in \Fig{firstSFacc}) and at the last time they do so (vertical axes). In Illustris (left), gas with lower angular momentum at the first crossing preferentially gains more angular momentum between these two events, namely during its participation in the `galactic wind fountain', compared with gas starting out with higher values. In No-Feedback (middle), there is a mild tendency to lose angular momentum between the two events, which however, in the absence of winds, do not represent a galactic fountain but instead `dynamical ejections'.}
\vspace{0.3cm}
\label{f:lastSFfirstSF}
\end{figure*}

Before that, we discuss No-Feedback during the \Fig{lastSFfirstSF} interval. In contrast to Illustris, it shows some overall {\it loss} of $0.1\dex$ in both mean magnitude and the magnitude of the vector sum. It is important to remember that in No-Feedback, about half of the tracers actually never leave the star-forming phase after they join it, hence the `first' and `last' times they do so are in fact the same event. In \Fig{lastSFfirstSF} we do not show these tracers, which by definition would lie on the 1:1 line. For the No-Feedback tracers for which these are indeed distinct events, the physical reason is very different from the typical case in Illustris. In No-Feedback, a tracer may leave the star-forming phase primarily for `dynamical' reasons, which occur naturally in the simulation and are not imposed as part of the sub-grid physics as is the case for wind ejections in Illustris. These dynamical reasons include for example tidal ejections during galaxy mergers and temporal density fluctuations around the star-formation density threshold due to weaker disturbances. The angular momentum loss occurring between these two events in No-Feedback is not negligible but is small compared to the losses incurred earlier and later, as shown in \Figs{NoFeed-firstSFacc}{NoFeed-laststarlastSF}, respectively.

Returning to Illustris, the period analyzed in \Fig{Ill2-lastSFfirstSF} includes both times when the tracer is in the star-forming phase and times in which it is outside of the galaxy in the `circum-galactic fountain'. The latter can be further broken down into times when the tracer belongs to a collisionless `wind particle' moving away from the galaxy, later times when it has recoupled to the normal gas phase and may be still moving away from the galaxy, and times when it is falling back toward the galaxy on its way to join the star-forming phase again. It is important to understand where the overall gains of $0.16\dex$ associated with this full period occur. \Fig{ejectionreturn} hence focuses on the last of possibly multiple `circum-galactic cycles' that each tracer goes through. The horizontal axis of each panel represents the same event: the time just before the last ejection into a wind (marked as the `later' (iii) in \Fig{cartoon}). The vertical axes show three subsequent events in chronological order: the very first snapshot after that same ejection event (\Fig{Ill2-ejectionbegin}); the maximum angular momentum the tracer has during that cycle through the halo before coming back to the galaxy (\Fig{Ill2-ejectionmax}); and the first snapshot after the tracer returns to the galaxy (namely either in the star-forming phase or directly as a star; \Fig{Ill2-ejectionreturn}).

\Fig{Ill2-ejectionbegin} shows that the angular momentum magnitudes of individual tracers increase between the two adjacent snapshots bracketing a wind ejection event. This is understandable, as a wind ejection implies an imposed momentum kick that increases the velocity of the tracer almost to the halo escape velocity. However, since these kicks are equally directed either `upwards' or `downwards' from the galaxy, they do not change the vector sum of the angular momentum of a population of ejected tracers. Indeed, we find that the vector sum between the two events shown in \Fig{Ill2-ejectionbegin} changes by only $0.04\dex$ (and in fact in the opposite direction, i.e.~it decreases). In other words, the momentum kicks associated with the wind ejection model itself do not change the overall angular momentum content of the tracers that enter the wind.

However during the time tracers spend in the circum-galactic medium following their ejection (\Fig{Ill2-ejectionmax}), they gain significant angular momentum. When each tracer is considered at the time its angular momentum magnitude is maximal between the ejection and the next time it appears in the central galaxy, the sum of magnitudes is $0.29\dex$ higher than it is right before the ejection, and the magnitude of the vector sum is $0.2\dex$ higher. The gains are particularly high for tracers that had lower angular momentum at the time they were ejected. Nevertheless, by the time tracers come back to the galaxy (\Fig{Ill2-ejectionreturn}), they return with angular momentum magnitudes that are on average essentially identical to those they had before the ejection (within the error on the mean, $\sim0.01\dex$, but some considerable spread), and a vector sum that is larger by $0.04\dex$.

In other words, a single cycle through the halo results in a net small degree of increased alignment between the tracers compared to the time before their ejection into the wind. Since in our galaxies of interest tracers go typically through several such cycles, this result ties well to the result discussed around \Fig{Ill2-lastSFfirstSF} that the full baryonic cycle in Illustris induces a $0.16\dex$ increase in net angular momentum.

Continuing forward in the tracers' evolution, \Fig{laststarlastSF} starts on the horizontal axes with the angular momentum at the last time tracers join the star-forming phase, and ends with the time they are converted to the stellar phase on the vertical axes. Namely, it pertains to evolution occurring within the galaxy, after all the evolution that occurs out in the halo. The two simulations again differ significantly. Illustris shows a tight correlation around the identity relation between the angular momenta at these times. In contrast, No-Feedback shows clear losses amounting to $0.14\dex$ for both the sum of magnitudes and the magnitude of the vector sum. Since the time tracers spend in the star-forming gas phase before forming stars is not very different between the two simulations, this result indicates that the presence of the galactic winds is changing the structure of galaxies in such a way that prevents the torques that exist otherwise and lead to angular momentum loss of the star-forming gas.

\begin{figure*}
\centering
\subfigure[]{
          \label{f:Ill2-ejectionbegin}
          \includegraphics[width=0.315\textwidth]{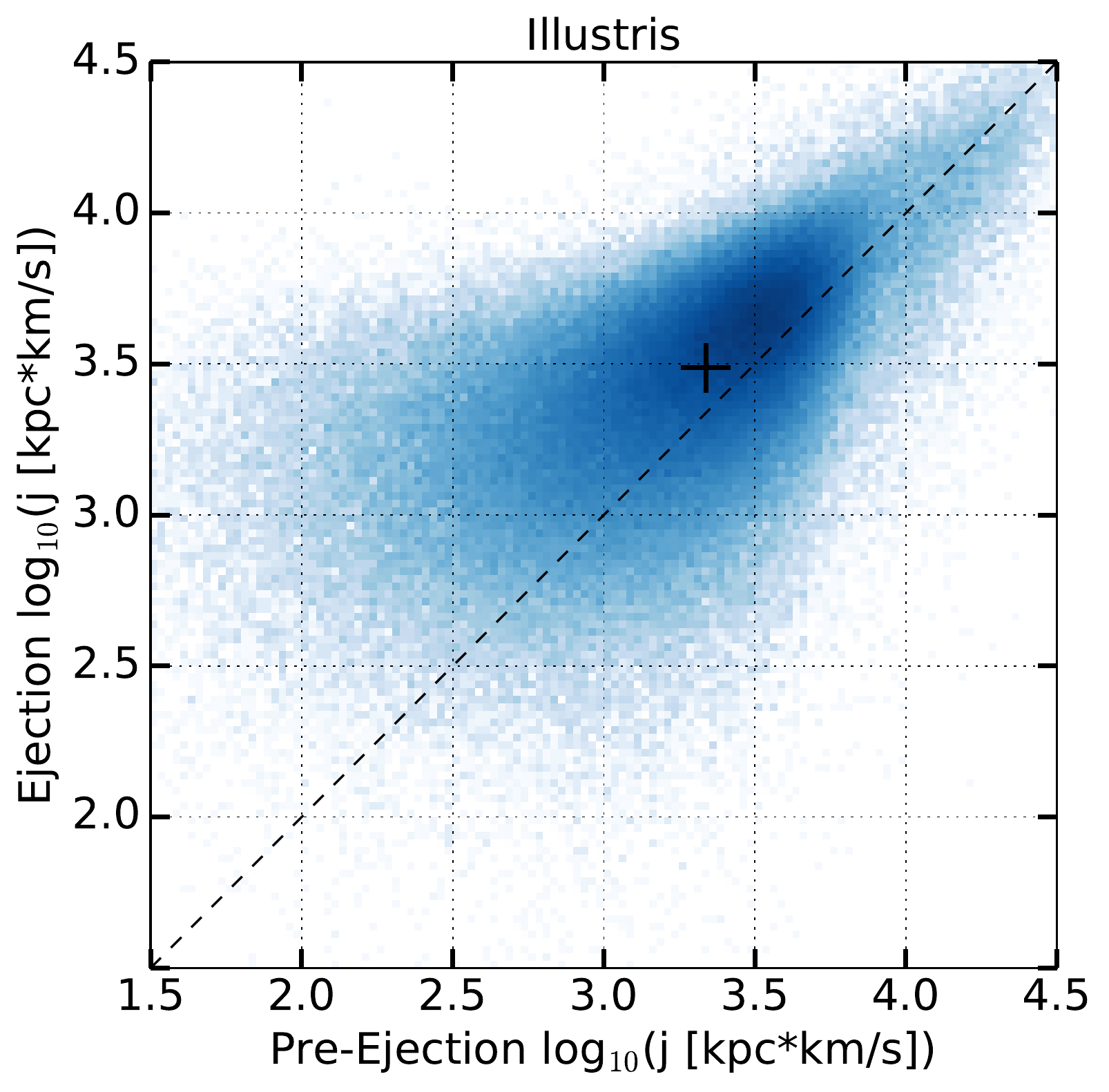}}
\subfigure[]{
          \label{f:Ill2-ejectionmax}
          \includegraphics[width=0.315\textwidth]{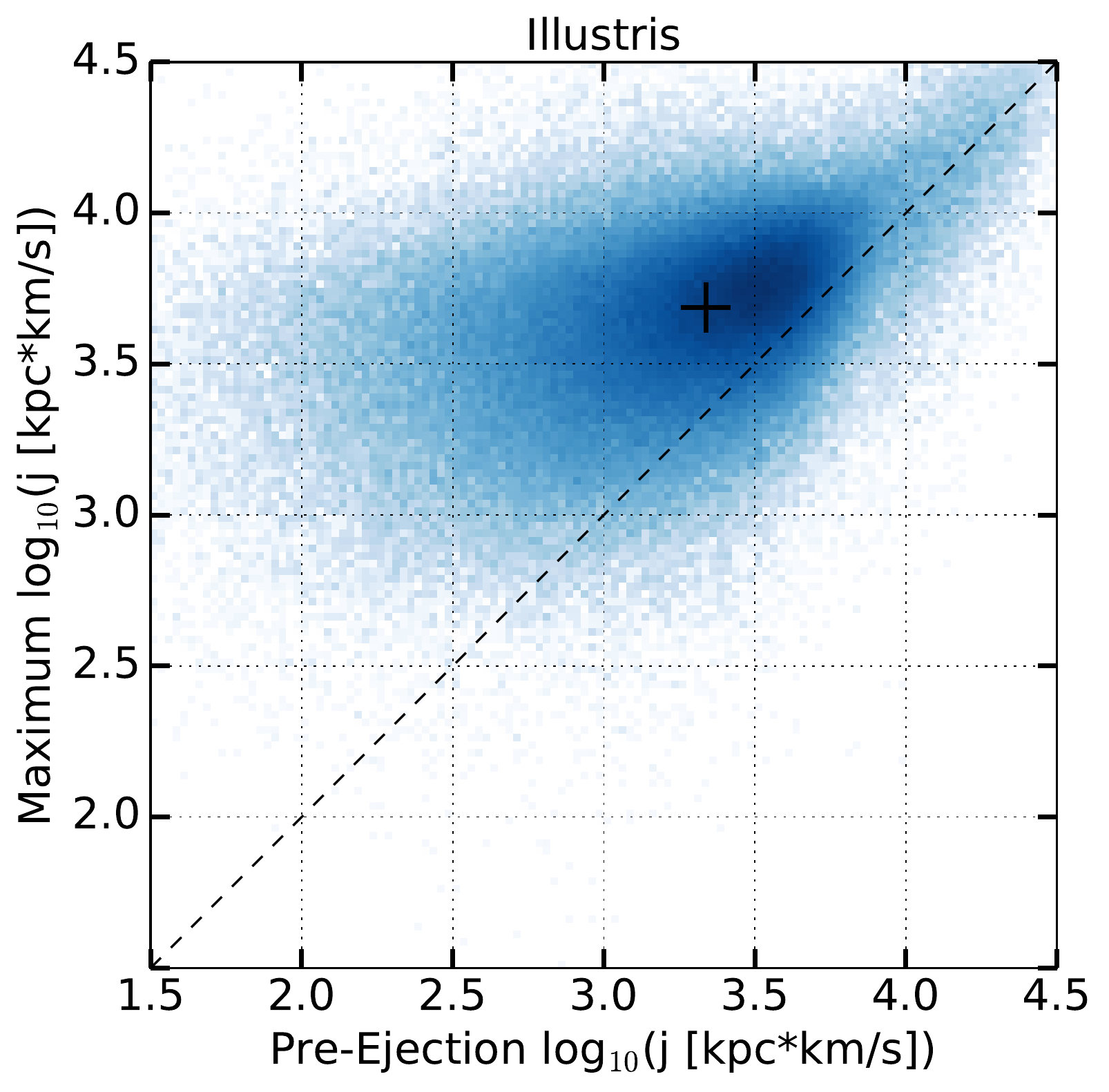}}
\subfigure[]{
          \label{f:Ill2-ejectionreturn}
          \includegraphics[width=0.351\textwidth]{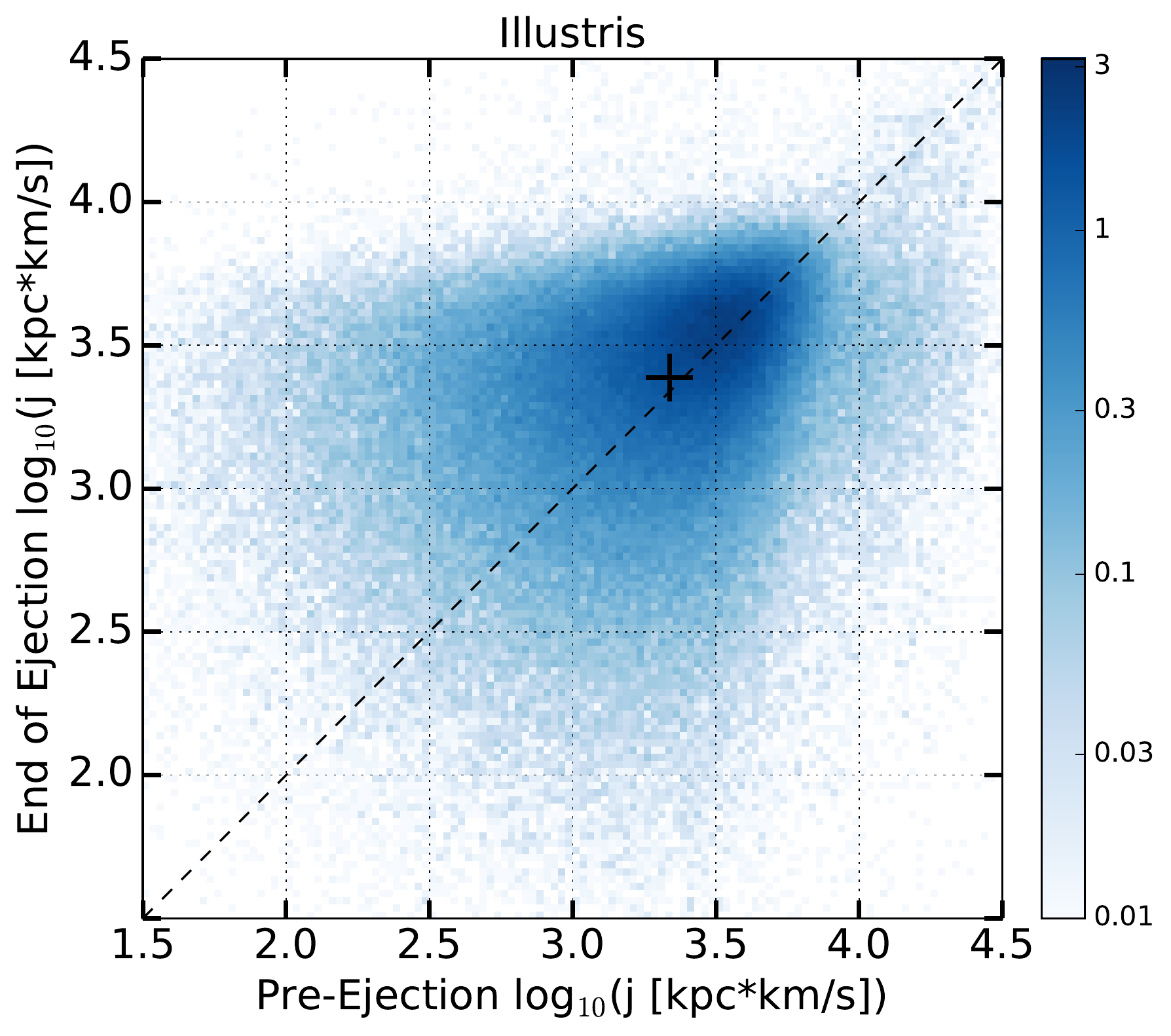}}
\caption{Joint probability distributions of angular momentum magnitudes of tracers in Illustris immediately before they are last recorded ejected in the wind (horizontal axes) and at three subsequent times on the vertical axes: immediately after that ejection (left); the end of the ejection, defined as the snapshot before coming back to the star-forming phase (right); and the time with the largest recorded angular momentum in between these two times (middle). During the ejection, tracers tend to gain angular momentum, especially those that have a low value before ejection, but by the time they return to the galaxy, the angular momentum largely returns to its pre-ejection value. The same holds when examining the distributions before, during, and after the {\it first} ejection, though the spread is larger.}
\vspace{0.3cm}
\label{f:ejectionreturn}
\end{figure*}

\begin{figure*}
\centering
\subfigure[]{
          \label{f:Ill2-laststarlastSF}
          \includegraphics[width=0.314\textwidth]{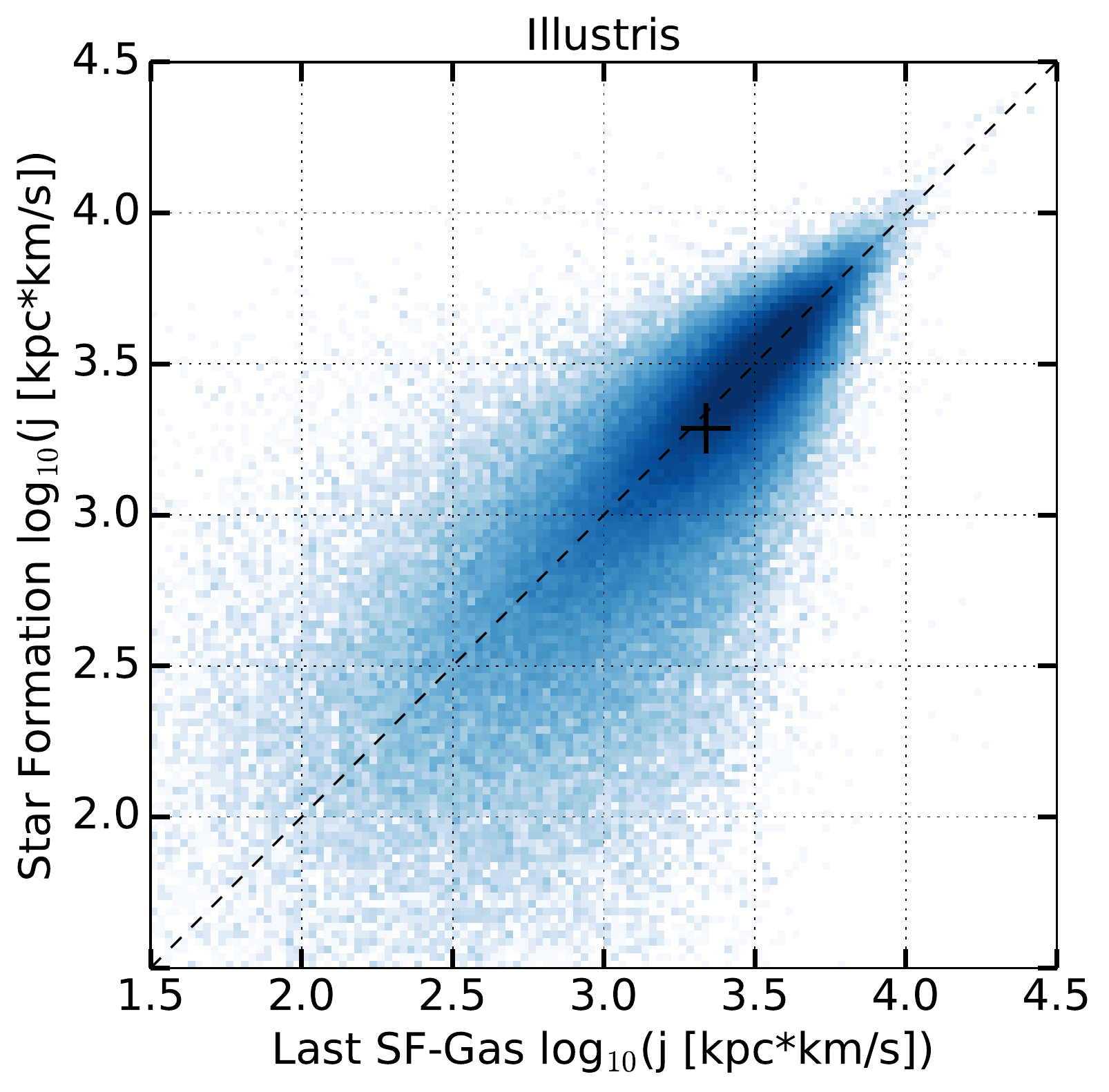}}
\subfigure[]{
          \label{f:NoFeed-laststarlastSF}
          \includegraphics[width=0.3215\textwidth]{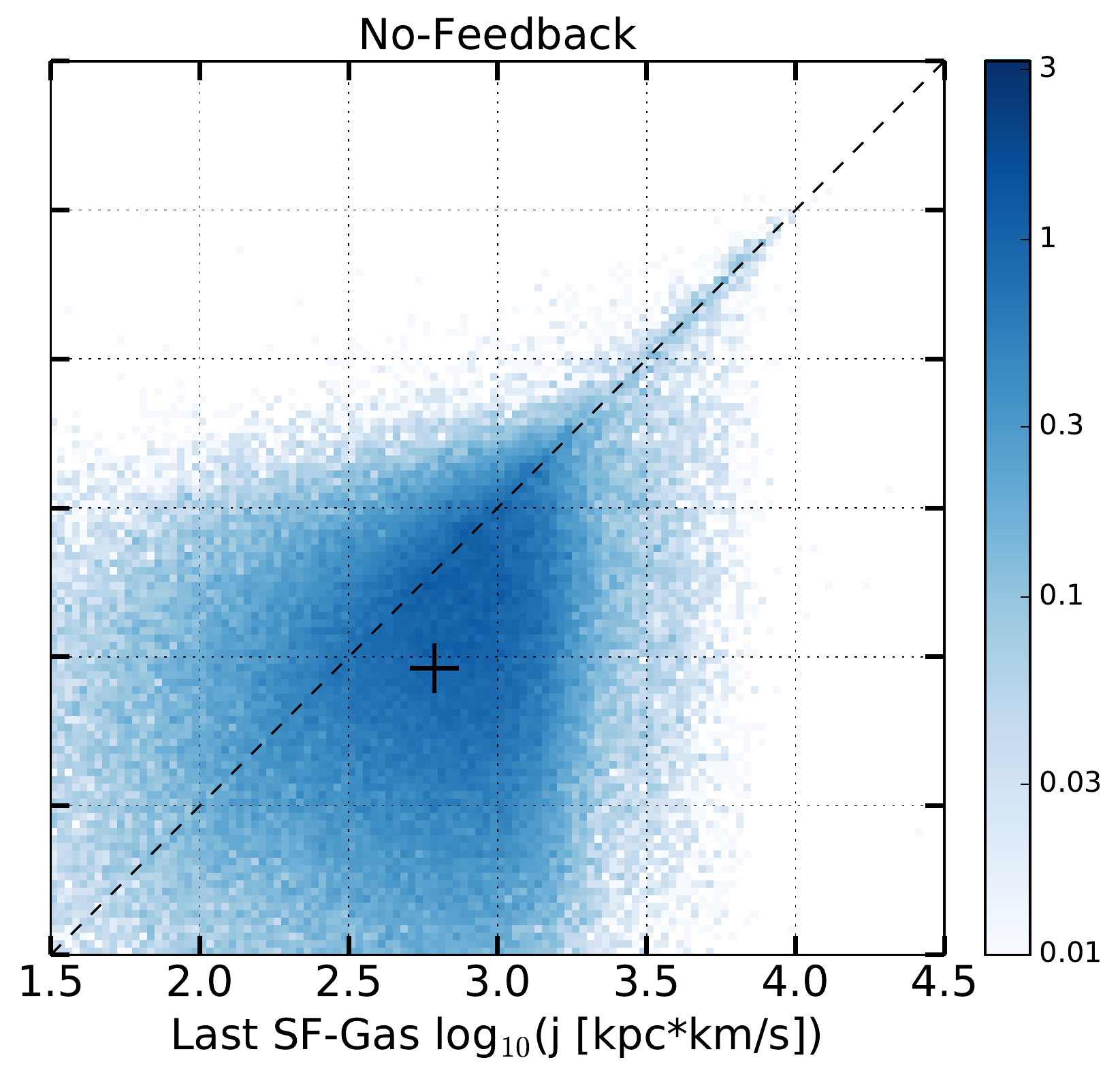}}
\subfigure[]{
          \label{f:laststarlastSF1D}
          \includegraphics[width=0.3355\textwidth]{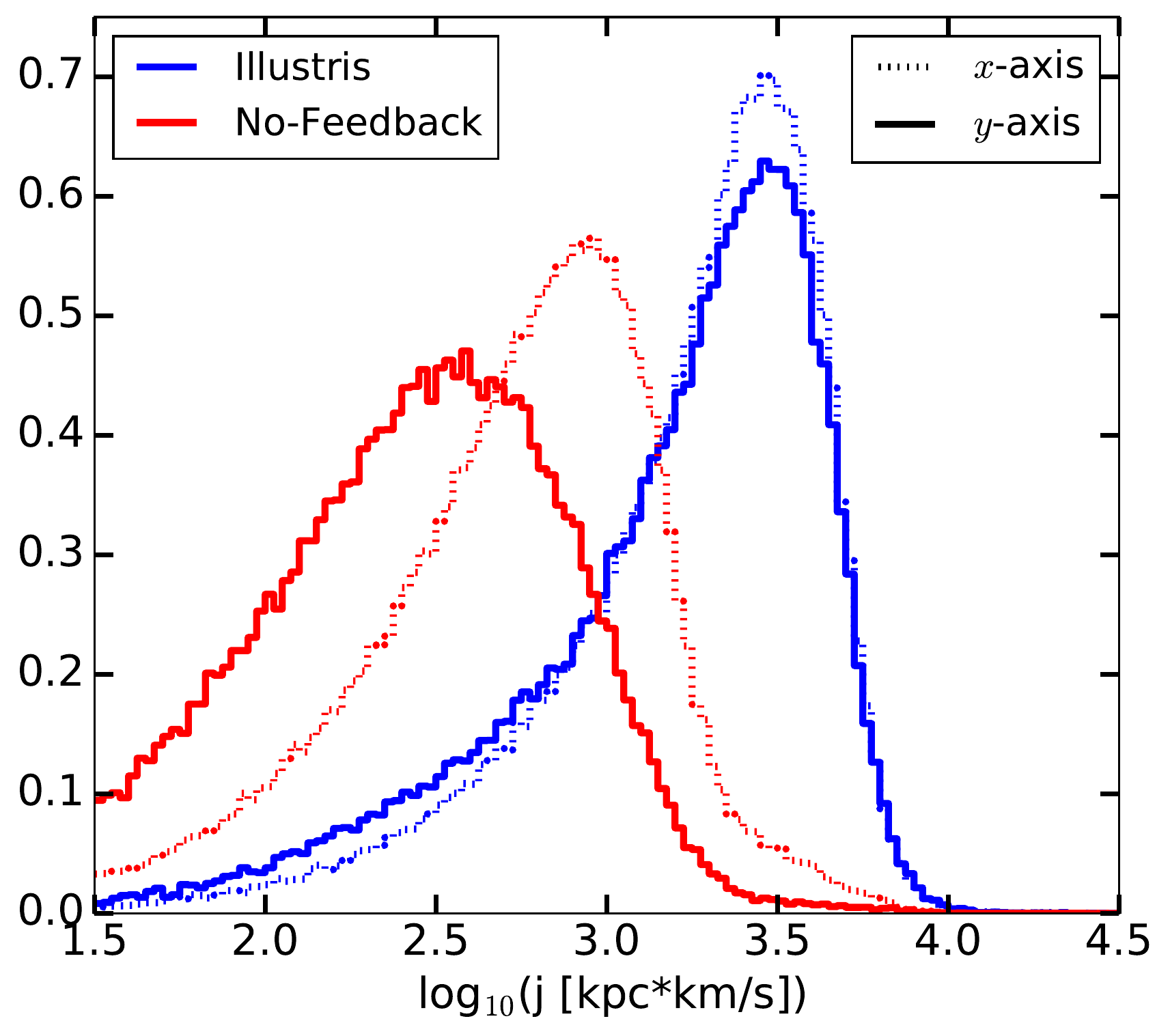}}
\caption{Joint and one-dimensional probability distributions of angular momentum magnitudes of tracers at the last time they cross the star-formation density threshold (horizontal axes; the same quantity as on the vertical axes in \Fig{lastSFfirstSF}) and at the last time they are converted from the gas phase to the stellar component (vertical axes). More so than any other period, these events in Illustris (left) are strongly correlated, while the tracers in No-Feedback still lose angular momentum by as much as $0.5\dex$ during their time in the star-forming phase before they are converted to stars.}
\vspace{0.3cm}
\label{f:laststarlastSF}
\end{figure*}

Finally in this sequence of events, we find the tightest correlation between the angular momentum tracers have at their time of star-formation and at $z=0$, shown in \Fig{redshift0laststar}. In Illustris, the stellar component experiences a minor gain of $0.03\dex$ in magnitude but essentially no change in the vector sum while in No-Feedback there is a small overall gain of $0.08\dex$ in both magnitude and vector sums.

\begin{figure*}
\centering
\subfigure[]{
          \label{f:Ill2-redshift0laststar}
          \includegraphics[width=0.314\textwidth]{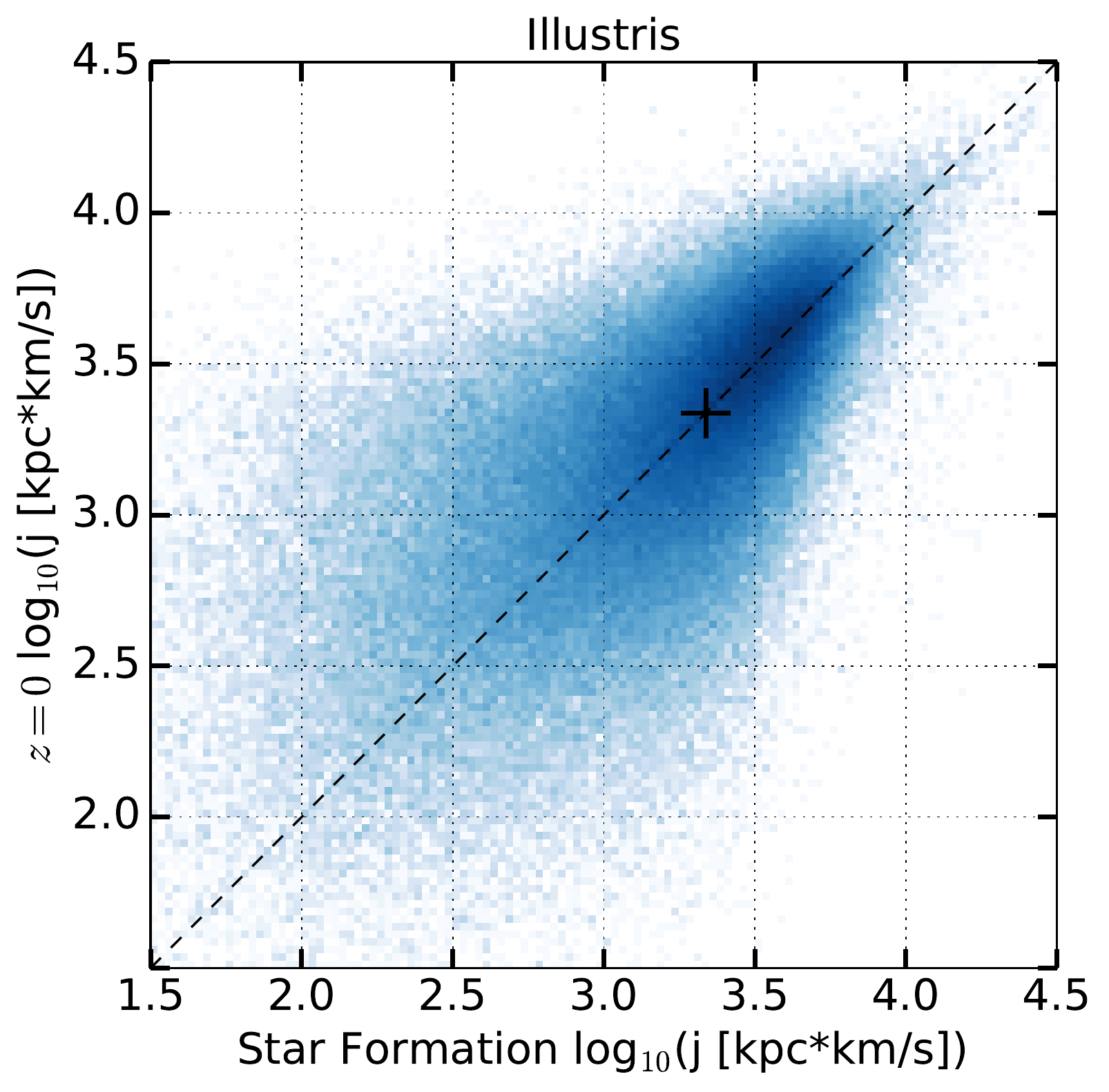}}
\subfigure[]{
          \label{f:NoFeed-redshift0laststar}
          \includegraphics[width=0.3215\textwidth]{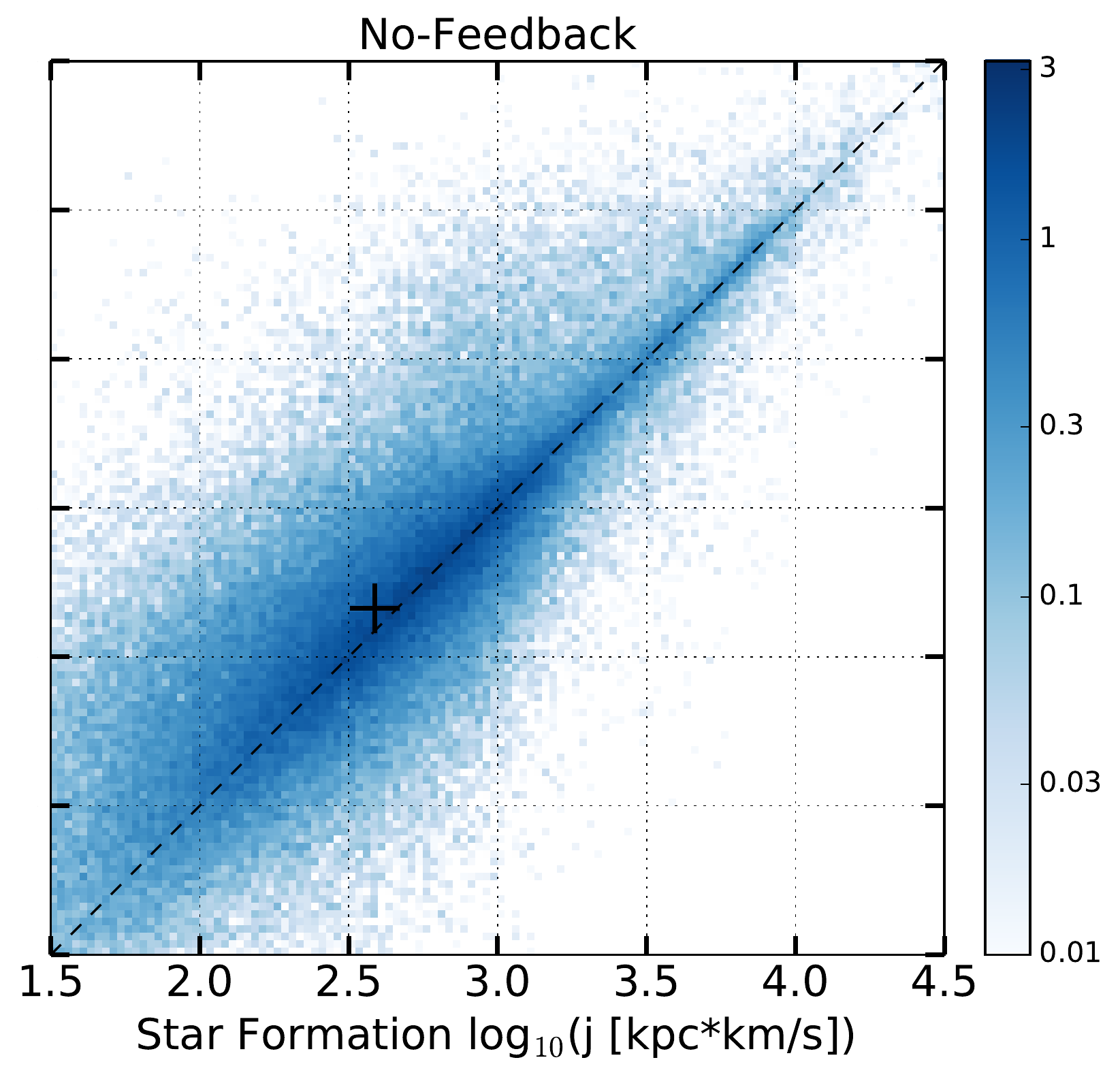}}
\subfigure[]{
          \label{f:redshiftlaststar1D}
          \includegraphics[width=0.3355\textwidth]{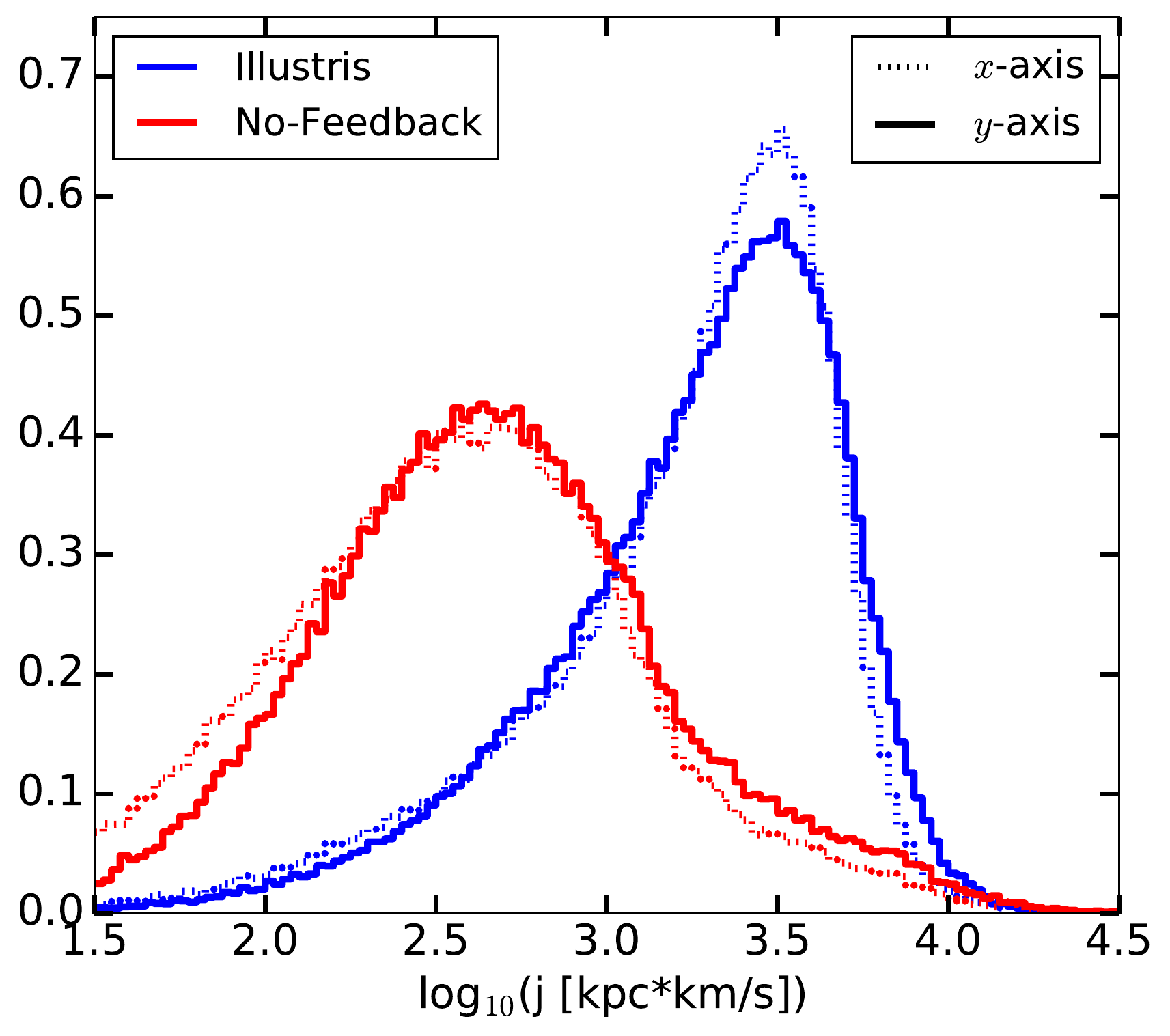}}
\caption{Joint and one-dimensional probability distributions of angular momentum magnitudes of tracers at the last time they are converted from the gas phase to the stellar component (horizontal axes; the same quantity as on the vertical axes in \Fig{laststarlastSF}) and as stars at $z=0$ (vertical axes). The two events are strongly correlated in both Illustris (left) and No-Feedback (middle), but the stellar component in No-Feedback experiences angular momentum gains of $0.08\dex$ compared to a negligible change in the stellar component of Illustris.}
\vspace{0.3cm}
\label{f:redshift0laststar}
\end{figure*}

\subsection{The angular momentum selection bias at halo accretion}
\label{s:results_bias}
After characterizing the angular momentum evolution inside halos, here we make several notes regarding the angular momentum differences between the simulations at the starting point of the preceding discussion, namely at the time baryons accrete onto the main progenitors of their $z=0$ host halos.

First, we make use of a third simulation, which we dub the `Feedback' simulation, that has identical initial conditions to No-Feedback, but the same subgrid models and parameter choices as Illustris. The tracers in this simulation can be compared on a one-to-one basis with the tracers in the No-Feedback simulation, as they have the same initial conditions. From this `Feedback' simulation, we select all $z=0$ stars from halos within the same mass range used for No-Feedback in the preceding analysis, identify those {\it same individual tracers} in No-Feedback, and compare their angular momentum at accretion between the two simulations\footnote{For direct comparison of stellar angular momentum, we exclude any of those tracers that are not stars by $z=0$ in No-Feedback, but this does not affect the numerical outcome.}. We find that both the vector and magnitude sums of the angular momentum at accretion of this identical set of tracers are equal between the two simulations. Namely, the addition of feedback does not modify the angular momentum value at halo accretion in Lagrangian space.

\begin{figure}
\centering
          \includegraphics[width=0.49\textwidth]{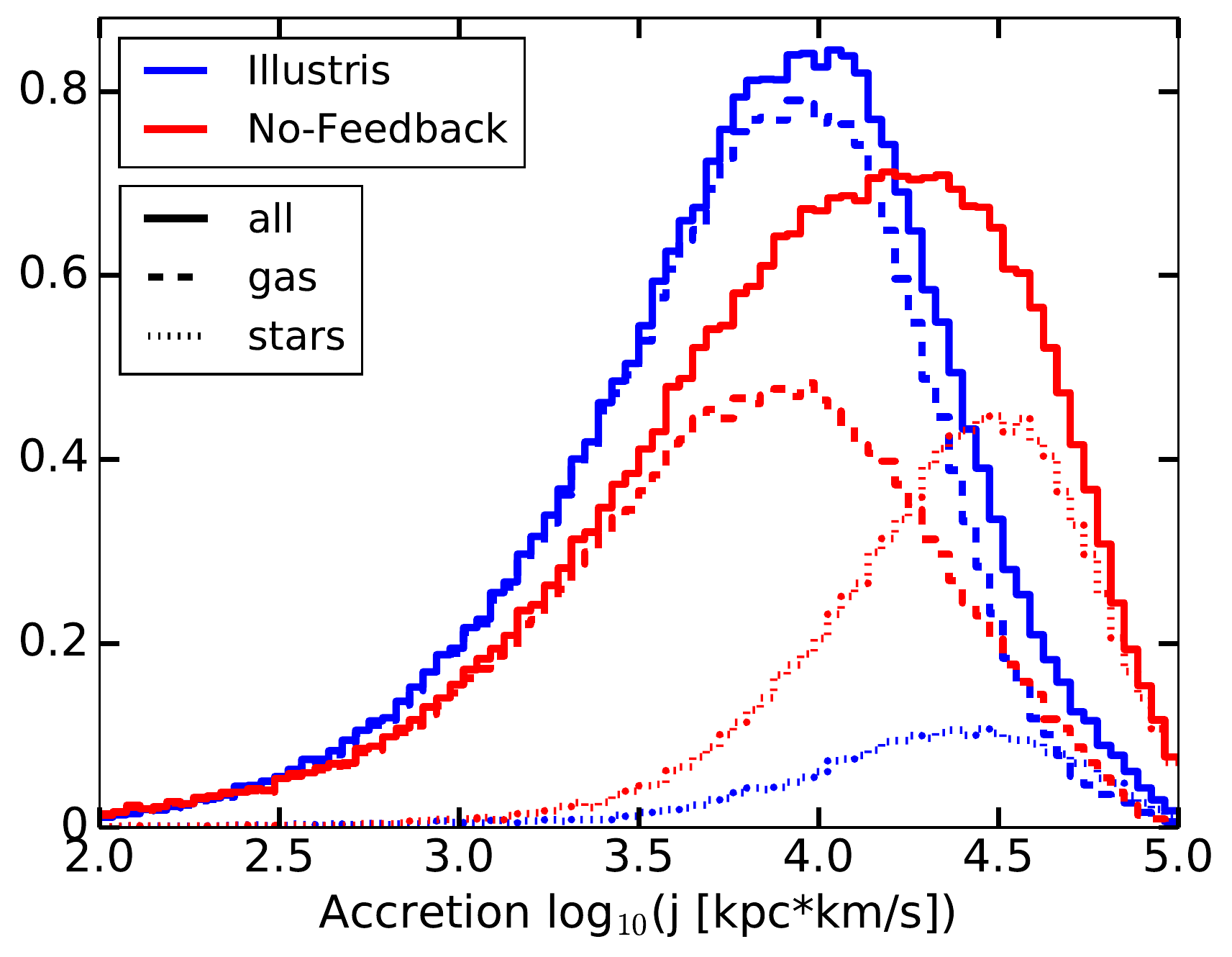}
\caption{Angular momentum magnitude probability density of tracers at their accretion time onto the main progenitor branch of their $z=0$ halo. For each simulation, the full tracer population that are in $z=0$ stars (solid) is split between those that were present at some point in the main galaxy as gas (dashed), and those that became a star already in a satellite galaxy and hence accreted to the main galaxy in stellar form (dotted). Stellar accretion typically has higher angular momentum, and it is far more significant in No-Feedback (red) than in Illustris (blue).}
\vspace{0.3cm}
\label{f:accretion}
\end{figure}

This contrasts with a comparison made when in each simulation tracers are selected independently as in Section \ref{s:analysis}. \Fig{accretion} shows the angular momentum distributions of $z=0$ stars at the time of accretion (solid), divided into the component that appears as gas in the main galaxy (dashed; the same as the distributions on the horizontal axes in \Figs{z0acc}{firstSFacc}) and the component that enters the main galaxy already as stars (dotted). While the gas-accreted distribution in No-Feedback (dashed red) is wider than the one in Illustris (dashed blue), they are peaked at the same value and have nearly identical magnitude sums. However, the vector sum of angular momentum in Illustris is $0.15\dex$ larger than that of No-Feedback, which reflects different degrees of self-alignment at accretion, as already noted in Section \ref{s:results_breakup} ($A_{\rm No-FB}^{\rm acc}=0.24$, $A_{\rm Ill}^{\rm acc}=0.36$).

These two results together imply that the galactic winds expel a fraction of the baryons and prevent them from becoming $z=0$ stars in a way that `selects' a more highly self-aligned set of gas tracers to end up as $z=0$ stars. This is done however without changing the angular momentum of those `selected' tracers at accretion.

Finally, the dotted curves in \Fig{accretion} represent material that forms stars in satellites and is accreted onto the main galaxy in stellar form (stellar mergers). We see that in both simulations, these stars accrete with a higher angular momentum\footnote{We also find (but do not show) that the accretion times of the different types of particles in \Fig{accretion} are unaffected by feedback: in both simulations stars are accreted later, in a similar way.}, having distributions that peak at $\log j \sim 4.5$. However, in Illustris this population only constitutes $\approx10\%$ of the $z=0$ stars while in No-Feedback it constitutes nearly half of them. The suppression of stellar accretion by feedback hence has a substantial effect on the overall angular momentum distribution that the baryons making up the $z=0$ galaxies have at their accretion time (solid). In this work we deliberately do not address the evolution of the stellar accretion inside the halo down to $z=0$, as it would require a distinctively different analysis from the gas component that is the focus of this work.

\section{Discussion and summary}
\label{s:summary}
Combining measurements of the angular momentum content of galactic disks with simple models that match galaxies to dark matter halos suggests that the specific angular momentum of galactic disks of different masses is very close (within $\approx20\%$) to the typical specific angular momentum of their host halo populations. The specific angular momentum content of a $z=0$ galaxy and its relation to that of its host halo can be considered using the following independent `bookkeeping' factors meant to separate physical effects:
\begin{enumerate}
\item The specific angular momentum content of dark matter accreted onto the halo, which integrated over cosmic time roughly gives the overall specific angular momentum of the $z=0$ dark matter halo.
\item The relation between the specific angular momentum of the baryons that accrete onto the halo along with the dark matter to that of the dark matter itself.
\item The (possible) specific angular momentum bias between all the baryons ever accreted onto the halo and the subset that end up in the galaxy.
\item The angular momentum evolution of those baryons that end up in the galaxy between the time they were accreted onto the halo and $z=0$.
\end{enumerate}
As we now discuss, this work has bearing for all of these steps except the second one, and in particular for the last two steps, which are shown here to be significantly affected by feedback processes.

First, if galaxy specific angular momentum was equal to that of the halo and all other factors did not introduce any differences, one would expect high angular momentum galactic disks to reside in halos with spins that are themselves higher than average, by a magnitude on the order of the standard deviation of the halo spin distribution, $\approx0.2-0.25\dex$ (e.g.~\citealp{BettP_07a}). This spread by itself would not suffice to explain the full range of specific angular momentum values of observed galaxies \citep{RomanowskyA_12a}, but it does need to be taken into account when comparing a subset of the galaxy population, namely galactic disks, to the full population of halos. Indeed, several recent cosmological hydrodynamical simulations found correlations between the angular momentum of halos and the galaxies they host \citep{TekluA_15a,Rodriguez-GomezV_17a,GrandR_17a}. We find a closely related trend here by the fact that our main analysis is based on the galaxies at the top $25\%$ of the specific angular momentum distribution, for which we find that at accretion, the vector sum is increased by $0.1\dex$ ($0.18\dex$) in Illustris (No-Feedback) with respect to the case of considering the full galaxy population. In other words, by selecting galaxies with high $z=0$ specific angular momentum, we select host halos that accrete (at least baryons) with higher angular momentum than typical. This shift has bearings for the overall picture in that a comparison of the angular momentum of galaxies at the top of the distribution to the typical angular momentum of halos includes a `halo selection bias'\footnote{The choice of whether all or just the top $25\%$ of galaxies are included does not, however, affect our conclusions regarding the angular momentum histories of baryons inside halos. Specifically, in Illustris the loss in the first interval (\Fig{firstSFacc}) is larger if the full population is selected ($0.44\dex$) compared to the case of our main analysis ($0.37\dex$), but the losses/gains in every subsequent interval remain unchanged.}.

Second, recent work suggests that the specific angular momentum of baryons at the time they accrete onto dark matter halos may be systematically higher than that of the dark matter accreted around the same time \citep{StewartK_13a,DanovichM_14a}, by up to $\approx0.2\dex$. This has to do with the higher quadrupole moment of cold gas in cosmic web streams. These conclusions were however drawn from a small number of `zoom-in' simulations and were mostly focused on $z\gtrsim1$, hence the quantitative significance of such an offset is not yet clear. In this work we have not examined the angular momentum of the dark matter itself and therefore do not show evidence to this effect or to the contrary, however it is important to keep this possibility in mind when considering the full picture.

Third, various effects can lead to a situation where the baryons that accrete over cosmic history onto the halo, with a distribution of angular momentum values, will not be sampled uniformly in angular momentum space in the galactic disks themselves. For example, gas accreted via cold streams, which has higher angular momentum at accretion, may be more likely to build the galaxy than hot gas accreted outside of streams \citep{StewartK_13a}. On the other hand, if galaxies are preferentially made of baryons that cool from the inner regions of their halos, the baryons making up the galaxies will be negatively biased in angular momentum relative to the full halo \citep{FallS_02a,KassinS_12a}. Another possibility, which is directly related to feedback and to the results in this work, is a bias generated by timing differences. Mass accreted at earlier cosmic times has lower angular momentum than mass accreted at later times (as in the classical tidal torque theory). Combined with the higher efficiency of galactic winds at ejecting gas out of galaxies at higher redshifts, this means that early-accreting, low-angular momentum gas can be biased against making up the final $z=0$ galaxy with respect to late-accreting, high-angular momentum gas \citep{BinneyJ_01a,BrookC_10a}. In this work we do not directly compare the baryons that do not make it to comprising the final $z=0$ galaxy to those that do, but we do show that the Illustris feedback does not significantly change the accretion time distribution of $z=0$ stars, and also has a weak effect on the angular momentum of individual tracers at accretion. We find however that the total angular momentum vector at accretion of $z=0$ stars that are accreted as gas is $0.15\dex$ higher in Illustris compared with No-Feedback, which requires further research. In addition, the fraction of $z=0$ stars that were formed in satellites (`ex-situ stars'), which are accreted with high angular momentum, is {\it suppressed} by the galactic winds in Illustris, thereby generating a bias at accretion that has an {\it opposite} sign to the overall difference between the two simulations, giving an `advantage' at accretion to the No-Feedback simulation. More research is needed to understand why the stellar accretion has higher angular momentum at accretion compared to gas that forms stars in-situ. One possibility is that it is related to the distinction between satellite and smooth accretion.

Finally, there are a variety of processes that may give rise to a situation where the baryons comprising the stars in a $z=0$ galaxy do not have the same angular momentum as they did when they entered the halo. Quantification of this scenario is the main focus of this work. We divide the time period between accretion into the halo and $z=0$ into several segments and reach the following findings, which are visually summarized in \Fig{summary}.
\begin{itemize}
\item Between accretion onto the halo and reaching the galaxy itself, we find that in No-Feedback baryons lose $0.49\dex$ and in Illustris they lose $0.37\dex$. Several processes probably operate during this period. Mutual torques between the dark matter and the gas due to their different spatial distributions can lead to angular momentum exchange from the former to the latter that results in the gas having an {\it increase} of $\approx0.1\dex$ in specific angular momentum compared to the dark matter, as shown by \citet{ZjupaJ_16a} using an {\it adiabatic} cosmological simulation. On the other hand, in the realistic case when radiative cooling is included, mass accreted via satellites (both as gas and as stars) is expected to experience dynamical friction that deprives it of orbital angular momentum. Torques in the inner part of the halo between the galaxy itself, the inflowing gas and the hot halo gas can significantly change the original angular momentum of all components \citep{RoskarR_10a,DanovichM_14a}. Our results suggest that feedback has a minimal effect on these processes, at least in a combined sense.
\item During the galactic fountain, namely between the first time baryons become part of the galactic star-forming gas and the last time they do so, we find gains of $\approx0.2\dex$ in the Illustris simulation. There is no true parallel to this time segment in the no-feedback simulation, since there are no galactic winds in that case. These gains in Illustris occur, in particular, to gas that initially has low angular momentum gas, as already seen in a handful of zoomed-in halos \citep{BrookC_12a,UeblerH_14a,ChristensenC_16a} even though, unlike those studies, the winds in Illustris are decoupled from the hydrodynamics. This suggests that it is not the kick itself that imparts lasting angular momentum gains, but rather several other processes likely operating during this period that do so. Gas that is ejected into the galactic wind spends of order the halo dynamical time at distances that are typically of order half of the virial radius. During this time its angular momentum can be enhanced by both large-scale tidal torques and local angular momentum exchange with the ambient halo may also occur via both gravitational and gas pressure forces.
\item Between reaching the galaxy itself (crossing the star-formation density threshold) for the last time and the actual star-formation time, we find that gas in Illustris on average does not lose any angular momentum, while in No-Feedback it loses on average as much as $0.14\dex$. Several processes probably operate during this period. Various types of non-axisymmetric distributions, such as spiral features, bars, and clumps formed by dynamical instabilities inside galactic disks, can cause angular momentum to flow out and mass to flow in. Dynamical interactions during galaxy mergers can also induce significant angular momentum losses. These interactions are expected to be stronger in No-Feedback because the low-mass galaxies that participate in mergers have much higher densities than those in Illustris, and indeed, our results show that these processes are strongly suppressed in the presence of galactic winds.
\item Of all the events we consider, the smallest changes in angular momentum content occur in the stellar phase, namely between the star-formation time and $z=0$. In Illustris, the stellar phase changes its angular momentum by $z=0$ by less than $\approx0.03\dex$. This is expected theoretically to be the case in the absence of bars, as the stellar component is dynamically hotter and non-dissipative (e.g.~\citealp{SellwoodJ_14a}). Our simulations are likely suppressing bar formation due to their limited resolution. Regardless, even in the presence of bars, where there is empirical indication for non-negligible secular evolution of angular momentum in disk galaxies (e.g.~\citealp{FoyleK_10a}), angular momentum exchange from the stellar component to the dark matter is expected to be inefficient \citep{ValenzuelaO_03a}. In the No-Feedback case, there are actually small gains at a level of $0.08\dex$. One possibility is that these stellar gains are obtained at the expense of the losses of gas component \citep{BournaudF_05b}, which are indeed stronger in No-Feedback.
\end{itemize}

Numerical work has shown in recent years that each of the `bookkeeping' steps discussed in the beginning of this section potentially involves  numerical factors with significant deviations from unity. Our results are qualitatively consistent with that work discussed in Section \ref{s:intro}, but demonstrating a true robustness to the hydrodynamics solver and input physics would require a direct code-to-code analysis. We show here that in the Illustris simulation, which reproduces the observed angular momentum of disk galaxies in $10^{12}\Msun$ halos, the last of these bookkeeping items, namely the angular momentum evolution of baryons inside halos, is composed of losses and gains of different magnitudes. The overall result of an offset of $\sim20\%$ between the angular momentum of galactic disks and of typical dark matter halos \citep{FallS_13a} is hence composed of a handful of numerical factors of $\approx0.1-0.4\dex$ each, which have distinct natures and origins. Some of these act to increase the baryonic angular momentum with respect to the dark matter one, and some in the opposite direction. It therefore remains a pressing theoretical challenge to understand the underlying reason for which they `conspire' to the simple and useful result of approximate `angular momentum retention' in galactic disks.

Future work will be required to clarify the physical processes and their relative roles in setting the angular momentum of disks: in other words, the `How?' and `Why?' presented in the Introduction. In particular it would be interesting to further investigate the lack of angular momentum loss of the star-forming gas phase in the Illustris simulation by better understanding the disk dynamics. Also important is to understand why and how (whether hydrodynamically or gravitationally) the low angular momentum gas gains angular momentum during wind ejections out in the halo (and what are the important effectors of torques and on which scales), while high angular momentum gas roughly maintains its angular momentum. In addition to new types of analysis, this will require smaller separation between snapshots than the full Illustris simulations currently provide. Additionally, it is necessary to further investigate the origins of self-alignment of the different baryonic components before accretion. The quantitative analysis of 'When', 'Where', and 'How much' presented here will serve as a starting point and guidance to these future studies.

\acknowledgements
We thank Rachel Somerville, Ari Maller and Avishai Dekel for useful discussions, as well as Vicente Rodriguez-Gomez for generating merger trees. We also thank the anonymous referee for their helpful comments. Analysis of the simulations was performed on the Odyssey cluster supported by the FAS Science Division Research Computing Group at Harvard University. The Flatiron Institute is supported by the Simons Foundation. SG acknowledges support provided by NASA through Hubble Fellowship grant HST-HF2-51341.001-A awarded by the STScI, which is operated by the Association of Universities for Research in Astronomy, Inc., for NASA, under contract NAS5-26555. GB acknowledges financial support from NASA grant NNX15AB20G and NSF grants AST-1312888 and AST-1615955.

\appendix
\label{s:appendix}
To further understand angular momentum differences between Illustris and No-Feedback, we examine the redshift distributions of the `events' identified in Section \ref{s:methods}, which are shown in \Fig{redshiftdist}. Unsurprisingly, feedback delays the formation of stars in Illustris (blue) and introduces a significant time lag between the first (yellow) and last (green) crossings of the star-formation density threshold. However, the distribution of halo accretion redshifts (red) is largely unaffected, indicating that while feedback changes the amount of time accreted gas spends in the halo before forming stars, it does not change the time at which gas is accreted in the first place.
\vspace{0.3cm}

\renewcommand{\thefigure}{A\arabic{figure}}
\setcounter{figure}{0}

\begin{figure*}
\centering
\subfigure[]{
          \label{f:Ill2-redshiftdist}          
          \includegraphics[width=0.49\textwidth]{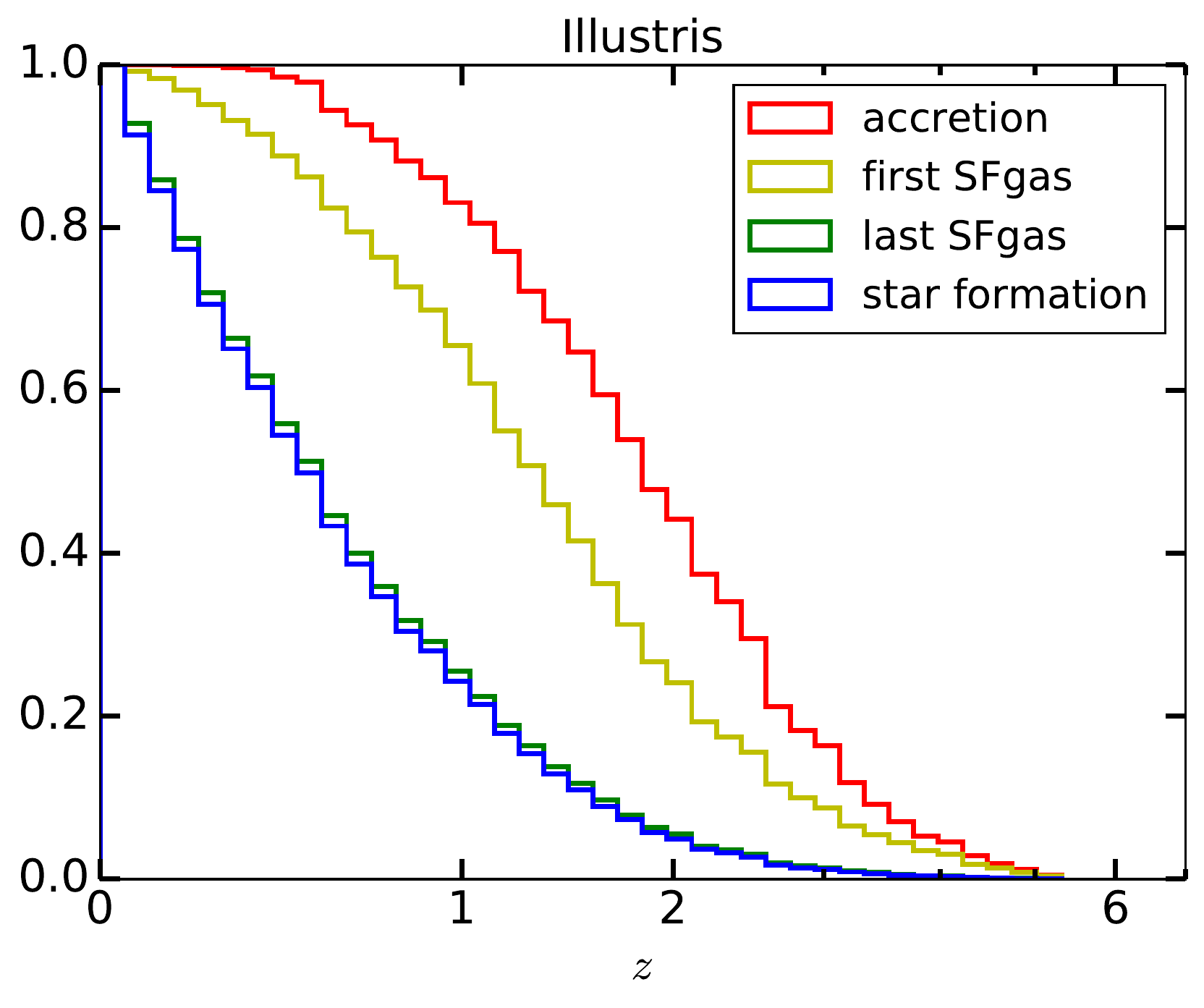}}
\subfigure[]{
          \label{f:NoFeed-redshitdist}
          \includegraphics[width=0.49\textwidth]{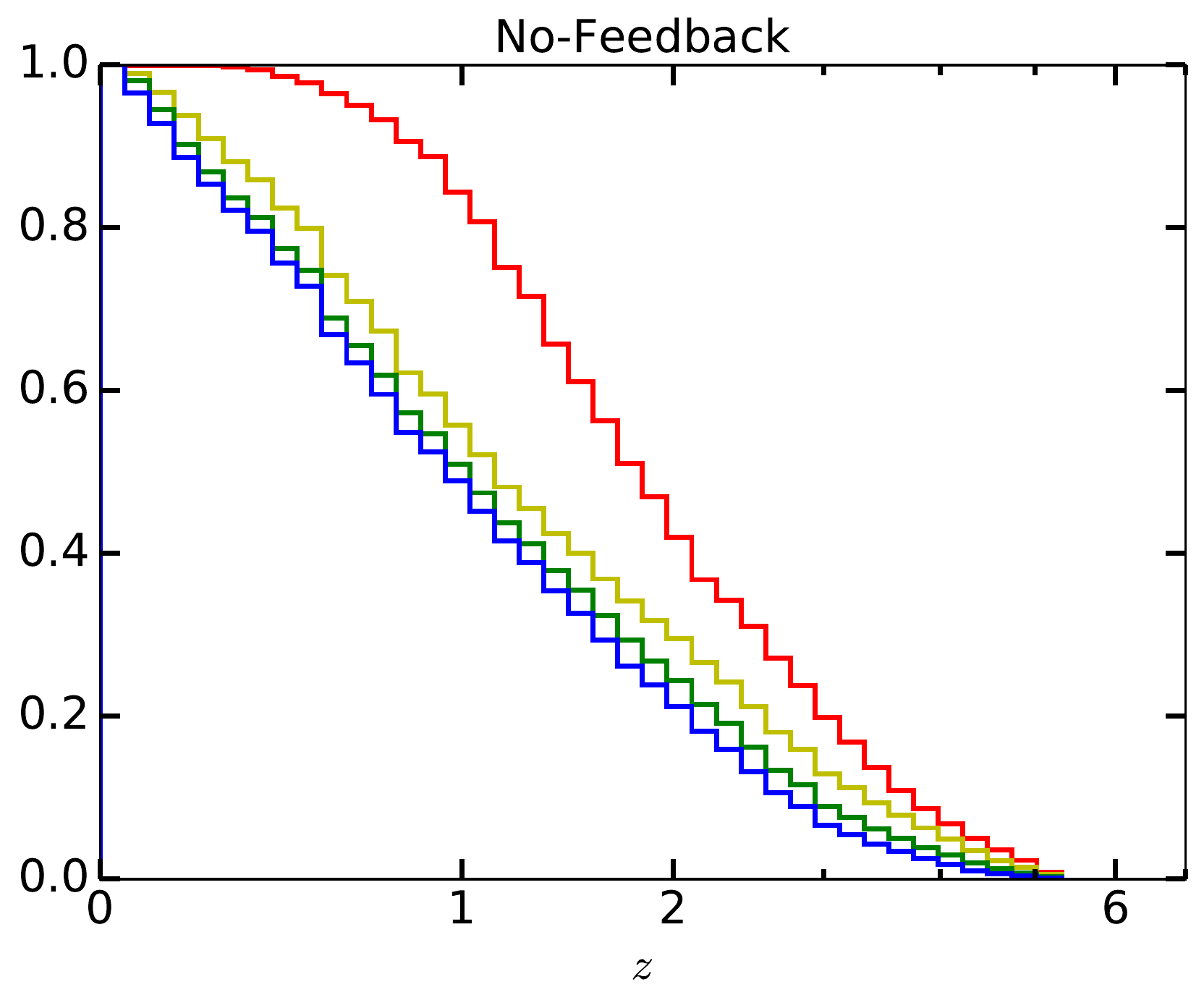}}
\caption{Cumulative redshift distributions of various events in Illustris (left) and No-Feedback (right) for $z=0$ stars that accreted onto the main galaxy as gas (i.e., ignoring stellar accretion). The addition of feedback delays star formation but leaves the halo accretion times essentially unaffected.}
\vspace{0.3cm}
\label{f:redshiftdist}
\end{figure*}

\bibliographystyle{apj}
\bibliography{references}

\label{lastpage}

\end{document}